\documentclass[preprint,10pt]{elsarticle}
\usepackage{soul}
\usepackage[colorlinks=true,linkcolor=black, citecolor=blue, urlcolor=blue]{hyperref}
\usepackage[usenames,dvipsnames,table]{xcolor}
\usepackage{colortbl}
\usepackage{url}
\usepackage{graphicx}
\usepackage{todonotes}
\usepackage[utf8]{inputenc}
\usepackage[T1]{fontenc}
\usepackage{lmodern} 
\usepackage{paralist}
\usepackage[inline]{enumitem}
\usepackage{caption}
\usepackage{ltablex}	
\usepackage{dingbat}
\definecolor{shadecolor}{gray}{0.95}
\usepackage{booktabs}
\usepackage{dblfloatfix}
\usepackage{pdflscape}
\usepackage{afterpage}
\usepackage{listings}
\usepackage{booktabs}
\usepackage{pifont}
\usepackage{lipsum}
\usepackage{multirow}
\usepackage[multiple]{footmisc}
\usepackage{hhline}
\usepackage{microtype}
\usepackage{array}
\biboptions{authoryear}
\usepackage{amssymb}
\usepackage{amsthm}
\usepackage{fourier-orns}
\usepackage{amsmath}
\usepackage{amsfonts}
\usepackage{stmaryrd}
\usepackage{amssymb}
\usepackage{enumitem}
\usepackage{tikz}
\newcommand*\circled[1]{\tikz[baseline=(char.base)]{
            \node[shape=circle,draw,inner sep=2pt] (char) {#1};}}

\usepackage{algorithm}
\usepackage{algpseudocode}
\algdef{SE}[DOWHILE]{Do}{doWhile}{\algorithmicdo}[1]{\algorithmicwhile\ #1}%

\usepackage{mathtools}
\makeatletter
\newcommand\Label[1]{&\refstepcounter{equation}(\theequation)\ltx@label{#1}&}
\makeatother

\newtheorem{definition}{Definition}[section]

\usepackage{adjustbox}
\usepackage{array}

\newcommand{\ue}{\mathbin{\mathord{\cup}\mathord{=}}}

\newcolumntype{P}[1]{>{\raggedright\arraybackslash}p{#1}}

\newcolumntype{M}[1]{>{\centering\arraybackslash}m{#1}}
\newcolumntype{N}[1]{>{\arraybackslash}m{#1}}

\colorlet{punct}{red!60!black}
\definecolor{background}{HTML}{EEEEEE}
\definecolor{delim}{RGB}{20,105,176}
\colorlet{numb}{magenta!60!black}

\lstdefinelanguage{json}{
    basicstyle=\scriptsize\ttfamily,
    numberstyle=\scriptsize,
    stepnumber=1,
    numbersep=8pt,
    showstringspaces=false,
    breaklines=true,
    literate=
     *{0}{{{\color{numb}0}}}{1}
      {1}{{{\color{numb}1}}}{1}
      {2}{{{\color{numb}2}}}{1}
      {3}{{{\color{numb}3}}}{1}
      {4}{{{\color{numb}4}}}{1}
      {5}{{{\color{numb}5}}}{1}
      {6}{{{\color{numb}6}}}{1}
      {7}{{{\color{numb}7}}}{1}
      {8}{{{\color{numb}8}}}{1}
      {9}{{{\color{numb}9}}}{1}
      {:}{{{\color{punct}{:}}}}{1}
      {,}{{{\color{punct}{,}}}}{1}
      {\{}{{{\color{delim}{\{}}}}{1}
      {\}}{{{\color{delim}{\}}}}}{1}
      {[}{{{\color{delim}{[}}}}{1}
      {]}{{{\color{delim}{]}}}}{1},
}

\lstdefinelanguage{myXML}
{
  columns=fullflexible,
  showstringspaces=false,
  commentstyle=\color{gray}\upshape,
  basicstyle=\scriptsize\ttfamily,
  numberstyle=\scriptsize,
  stepnumber=1,
  numbersep=8pt,
  morestring=[b]",
  morestring=[s]{>}{<},
  morecomment=[s]{<?}{?>},
  stringstyle=\color{black},
  identifierstyle=\color{darkblue},
  keywordstyle=\color{cyan},
  morekeywords={xmlns,version,type}
}

\definecolor{dkgreen}{rgb}{0,0.6,0}
\definecolor{gray}{rgb}{0.5,0.5,0.5}
\definecolor{mauve}{rgb}{0.58,0,0.82}
\definecolor{darkblue}{rgb}{0.0,0.0,0.6}

\lstdefinelanguage{turtle}{
  frame=tb,
  language=xml,
  aboveskip=-2mm,
  belowskip=-2mm,
  showstringspaces=true,
  columns=flexible,
  basicstyle={\scriptsize\ttfamily\color{darkblue}},
  numbers=none,
  numberstyle=\tiny\color{gray},
  keywordstyle=\color{red},
  commentstyle=\color{dkgreen},
  stringstyle=\color{mauve},
  breaklines=true,
  breakatwhitespace=false,
  tabsize=1,
  morecomment=[s]{<!--}{-->},
  morecomment=[s]{\(.}{.\)},
  identifierstyle=\color{black}
} 

\makeatletter
\def\old@comma{,}
\catcode`\,=13
\def,{%
  \ifmmode%
    \old@comma\discretionary{}{}{}%
  \else%
    \old@comma%
  \fi%
}
\makeatother

\newcommand{\tick}{\ding{51}}

\algrenewcommand\algorithmicrequire{\textbf{Pre:}}
\algrenewcommand\algorithmicensure{\textbf{Post:}}
\algrenewcommand\algorithmicindent{1.0em}%

\def\Plus{\texttt{+}}
\newcommand\tab[1][.5cm]{\hspace*{#1}} 

\newcommand{\Glb}[1]{\textit{#1}}
\newcommand{\Src}[1]{\textsf{#1}}
\newcommand{\Rdf}[1]{\texttt{#1}}

\journal{}

\begin{document}

\begin{frontmatter}

\title{An Integration-Oriented Ontology to Govern Evolution in Big Data Ecosystems}

\author[upc]{Sergi Nadal}\corref{mycorrespondingauthor}
\cortext[mycorrespondingauthor]{Corresponding author. Address: Campus Nord Omega-125, UPC - dept ESSI, C/Jordi Girona Salgado 1-3, 08034 Barcelona,
Spain}
\ead{snadal@essi.upc.edu}

\author[upc]{Oscar Romero}
\ead{oromero@essi.upc.edu}

\author[upc]{Alberto Abelló}
\ead{aabello@essi.upc.edu}

\author[uoi]{Panos Vassiliadis}
\ead{pvassil@cs.uoi.gr}

\author[ulb]{Stijn Vansummeren}
\ead{stijn.vansummeren@ulb.ac.be}

\address[upc]{Universitat Polit{\`e}cnica de Catalunya - BarcelonaTech}
\address[uoi]{University of Ioannina}
\address[ulb]{Universit{\'e} Libre de Bruxelles}


\begin{abstract}
Big Data architectures allow to flexibly store and process heterogeneous data, from multiple sources, in their original format. The structure of those data, commonly supplied by means of REST APIs, is continuously evolving. Thus data analysts need to adapt their analytical processes after each API release. This gets more challenging when performing an integrated or historical analysis. To cope with such complexity, in this paper, we present the Big Data Integration ontology, the core construct to govern the data integration process under schema evolution by systematically annotating it with information regarding the schema of the sources. We present a query rewriting algorithm that, using the annotated ontology, converts queries posed over the ontology to queries over the sources. To cope with syntactic evolution in the sources, we present an algorithm that semi-automatically adapts the ontology upon new releases. This guarantees ontology-mediated queries to correctly retrieve data from the most recent schema version as well as correctness in historical queries. A functional and performance evaluation on real-world APIs is performed to validate our approach.
\end{abstract}

\end{frontmatter}

\section{Introduction} \label{sec:introduction}

Big Data ecosystems enable organizations to evolve their decision making processes from classic stationary data analysis \cite{DBLP:journals/jdwm/AbelloDEGMNPRTVV13} (e.g., transactional) to situational data analysis \cite{DBLP:conf/birte/LoserHM08} (e.g., social networks). Situational data are commonly obtained in the form of data streams supplied by third party data providers (e.g., Twitter or Facebook), by means of web services (or APIs). Those APIs offer a part of their data ecosystem at a certain price allowing external data analysts to enrich their data pipelines with them. With the rise of the RESTful architectural style for web services \cite{DBLP:conf/www/PautassoZL08}, providers have flexible mechanisms to share such data, usually semi-structured (i.e., JSON), over web protocols (e.g., HTTP). However, such flexibility can be often a disadvantage for analysts. In contrast to other protocols offering machine-readable contracts for the structure of the provided data (e.g., SOAP), web services using REST typically do not publish such information. Hence,  \emph{analysts need to go over the tedious task of carefully studying the documentation and adapting their processes to the particular schema provided}. Besides the aforementioned complexity imposed by REST APIs, there is a second challenge for data analysts. \emph{Data providers are constantly evolving such endpoints}\footnote{\url{https://dev.twitter.com/ads/overview/recent-changes}}\footnote{\url{https://developers.facebook.com/docs/apps/changelog}}, hence \emph{analysts need to continuously adapt the dependent processes to such changes}.
Previous work on schema evolution has focused on software obtaining data from relational views \cite{DBLP:journals/jodsn/ManousisVP15,DBLP:journals/is/SkoulisVZ15}. Such approaches rely on the capacity to veto changes affecting consumer applications. Those techniques are not valid in our setting, due to the lack of explicit schema information and the impossibility to prevent changes from third party data providers.

\emph{Given this setting, the problem is how to aid the data analyst in the presence of schema changes by (a) understanding what parts of the data structure change and (b) adapting her code to this change}.

Providing an integrated view over an evolving and heterogeneous set of data sources is a challenging problem, commonly referred as the data variety challenge \cite{DBLP:journals/internet/HorrocksGKW16}, that traditional data integration techniques fail to address. An approach to tackle it is to leverage on Semantic Web technologies, and the so-called ontology-based data access (OBDA). OBDA are a class of systems that enable end-users to query an integrated set of heterogeneous and disparate data sources decreasing the need for IT support \cite{DBLP:journals/jods/PoggiLCGLR08}. OBDA achieves its purpose by providing a conceptualization of the domain of interest, via an ontology, allowing users to pose ontology-mediated queries (OMQs), and thus creating a separation of concerns between the conceptual and the database level. Due to the simplicity and flexibility of ontologies, they constitute an ideal tool to model such heterogeneous environments. However, such flexibility is also one of its biggest drawbacks, as OBDA currently has no means to provide continuous adaptation to changes in the sources (e.g., schema evolution), and thus causing queries to crash.

The problem is not straightforwardly addressable, as current OBDA approaches, which are built upon generic reasoning in description logics (DLs), represent schema mappings following the \textit{global-as-view} (GAV) approach \cite{DBLP:conf/pods/Lenzerini02}. In GAV, elements of the ontology are characterized in terms of a query over the source schemata. This provides simplicity in the query answering methods, which consists of unfolding the queries to the sources. Changes in the source schemata, however, will invalidate the mappings. In contrast, \textit{local-as-view} (LAV) schema mappings characterize elements of the source schemata in terms of a query over the ontology. They are naturally suited to accomodate dynamic environments, as we will see. The trade-off however, comes at the expense of query answering, which becomes a computationally complex task that might require reasoning \cite{DBLP:journals/tlsdkcs/0001RA16}. To this end, we aim to bridge this gap by providing a new approach to OBDA with LAV mapping assertions, while maintaining query answering tractable. 
We follow a vocabulary-based approach which rely on tailored metadata models to design the ontology (i.e., a set of design guidelines). This allows to annotate the data integration constructs with semantic annotations, enabling to automate the process of evolution and resolve query answering without ambiguity. Oppositely to reasoning-based approaches, vocabulary-based OBDA is not limited by the expressiveness of a concrete DL for query answering, as it does not rely on generic reasoning techniques but on ad-hoc algorithms that leverage such semantic annotations.

Our approach builds upon the well-known framework for data integration \cite{DBLP:conf/pods/Lenzerini02}, and it is divided in two levels represented by graphs (i.e., Global and Source graphs) in order to provide analysts with an integrated and format-agnostic view of the sources. By relying on wrappers (from the well-known mediator/wrapper architecture for data integration \cite{DBLP:books/daglib/0020812}) we are able to accomodate different kinds of data sources, as the query complexity is delegated to wrappers and the ontology is only concerned with how to join them and what attributes are projected. Additionally, we allow the ontology to contain elements that do not exist in the sources (i.e., syntactic sugar for data analysts), such as taxonomies, to facilitate querying. The process of query answering is reduced to properly resolving the LAV mapping assertions, relying on the annotated ontology, in order to construct an expression fetching the attributes provided by the wrappers. Finally, we exploit this structure to handle the evolution of source schema via semi-automated transformations on the ontology upon service releases.

\paragraph{Contributions} The main contributions of this paper are as follows:
\begin{itemize}
  \item We introduce a structured ontology based on an RDF vocabulary that allows to model and integrate evolving data from multiple providers. As an add-on, we take advantage of RDF's nature to semantically annotate the data integration process.
    \item We provide a method that handles schema evolution on the sources. According to our industry applicability study, we flexibly accommodate source changes by only applying changes to the ontology, dismissing the need to change the analyst's queries.
    \item We present a query answering algorithm that using the annotated elements in the ontology is capable of unambiguously resolving LAV mappings. Given a OMQ over the ontology, we are capable of manipulating it yielding an equivalent query over the sources. We further provide a theoretical and practical study of its complexity and limitations.
  \item We assess our method by performing a functional and performance evaluation. The former reveals that our approach is capable of semi-automatically accomodating all structural changes concerning data ingestion, which on average makes up 71.62\% of the changes occurring on widely used APIs.
\end{itemize}

\paragraph{Outline} The rest of the paper is structured as follows. Section \ref{sec:overview} describes a running example and formalizes the problem at hand. Section \ref{sec:ontology} discusses the constructs of the Big Data Integration ontology and its RDF representation. Section \ref{sec:evolution} introduces the techniques to manage schema evolution. Section \ref{sec:query_answering} presents the query answering algorithm.  Section \ref{sec:experiments} reports on the evaluation results. Sections \ref{sec:related} and \ref{sec:conclusions} discuss related work and conclude the paper, respectively.
\section{Overview} \label{sec:overview}

\begin{figure}[h!]
\centering
\includegraphics[width=1\linewidth]{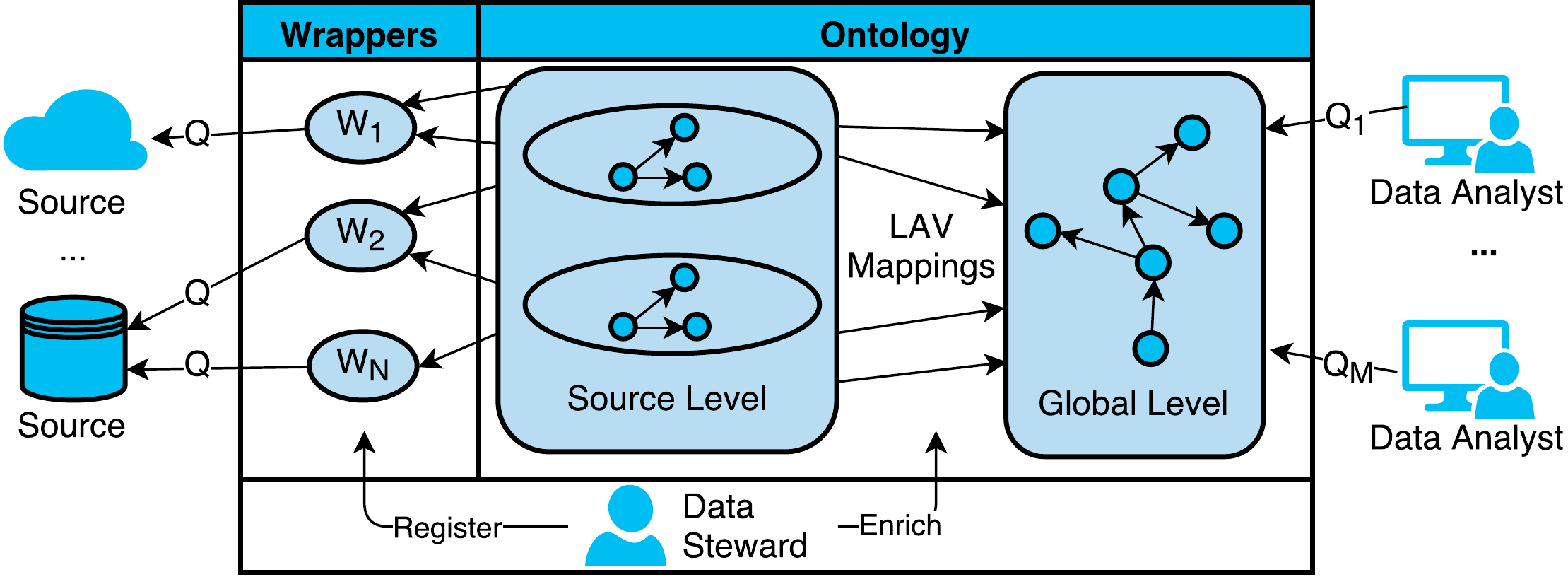}
\caption{High-level overview of our approach} \label{fig:big_picture}
\end{figure}

Our approach (see Figure \ref{fig:big_picture}) relies on a two-level ontology of RDF named graphs to accomodate schema evolution in the data sources. Such graphs are built based on a RDF vocabulary tailored for data integration. Precisely, we divide it into the \textit{Global graph} ($\mathcal{G}$), and the \textit{Source graph} ($\mathcal{S}$). Briefly, $\mathcal{G}$ represents an integrated view of the domain of interest (also known as domain ontology), while $\mathcal{S}$ represents data sources, wrappers and their schemata. On the one hand, data analysts issue OMQs 
to $\mathcal{G}$. We also assume a triplestore with a SPARQL endpoint supporting the RDFS entailment regime (e.g., subclass relations are automatically inferred) \cite{DBLP:conf/semweb/Horst04}. On the other hand, we have a set of data sources, each with a set of wrappers querying it. Different wrappers for a data source represent different schema versions. Under the assumption that wrappers provide a flat structure in first normal form, we can easily depict an accurate representation of their schema into $\mathcal{S}$. To acommodate a LAV approach, each wrapper in $\mathcal{S}$ is related to the fragment of $\mathcal{G}$ for which it provides data.

The management of such a complex structure (i.e., modifying it upon schema evolution in the sources) is a hard task to automate. To this end, we introduce the role of data steward as an analogy to the database administrator in traditional relational settings. Aided by semi-automatic techniques, s/he is responsible for (a) registering the wrappers of newly incoming, or evolved, data sources in $\mathcal{S}$, and (b) make such data available to analysts by defining LAV mappings to $\mathcal{G}$ (i.e., enriching the ontology with the mapping representations).
With such setting, intuitively the problem consists of given a query over $\mathcal{G}$, to derive an equivalent query over the wrappers leveraging on $\mathcal{S}$. Throughout the rest of this section, we introduce the running example and the formalism behind our approach. To make a clear distinction among concepts, hereinafter, we will use \Glb{italics} to refer to elements in $\mathcal{G}$, while \Src{sans serif} font to refer to elements in $\mathcal{S}$. Additionally, to refer to RDF constructs, we will use \Rdf{typewriter} font.

\subsection{Running Example} \label{sec:runningExample}

\begin{figure}[h!]
\centering
\includegraphics[width=0.75\linewidth]{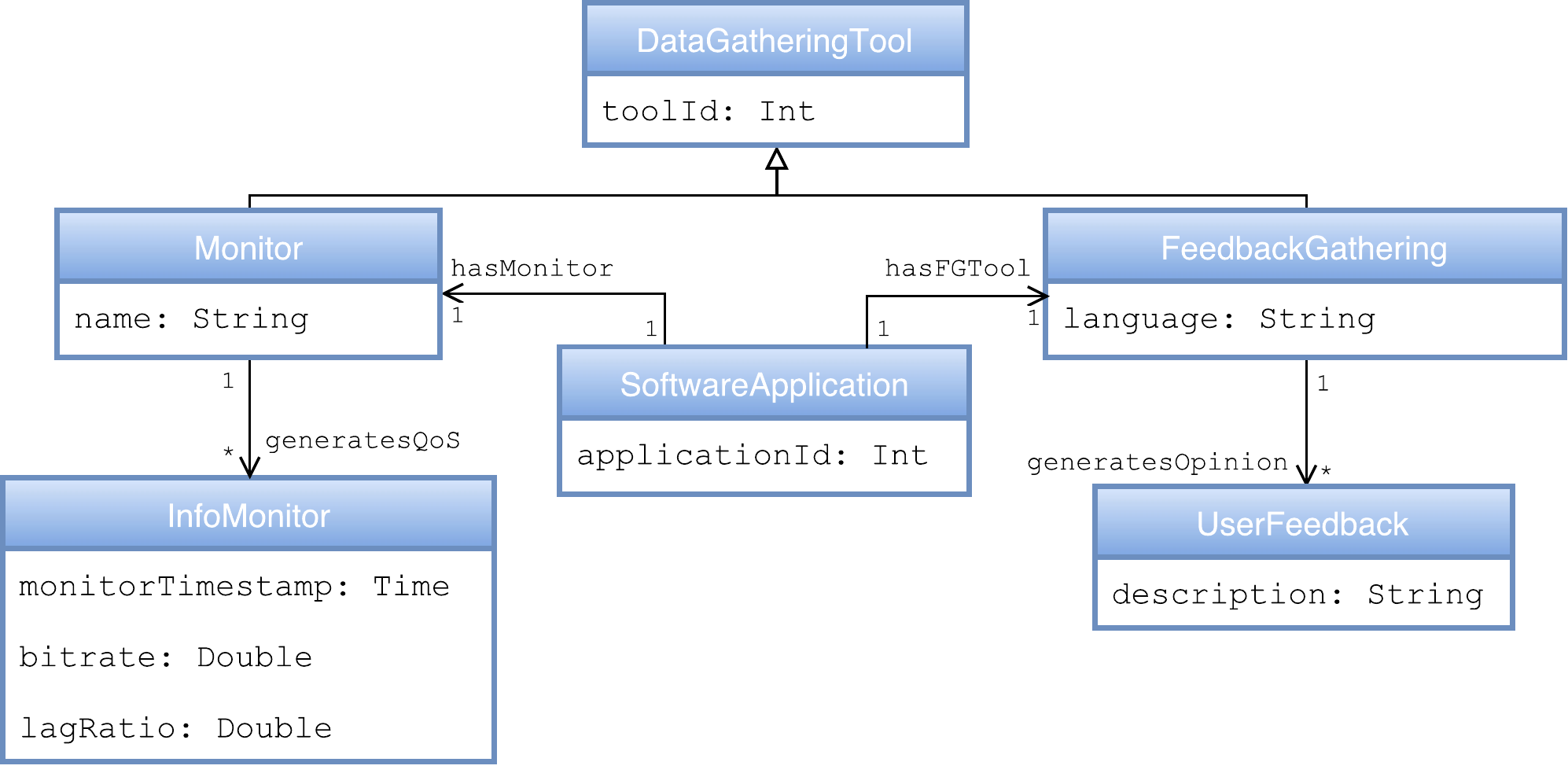}
\caption{UML conceptual model for the SUPERSEDE case study} \label{fig:UML}
\end{figure}

As an exemplary use case we take the H2020 SUPERSEDE project\footnote{\url{https://www.supersede.eu}}. It aims to support decision-making in the evolution and adaptation of software services and applications (i.e., \Glb{SoftwareApps}) by exploiting end-user feedback and monitored runtime data, with the overall goal of improving end-users’ quality of experience. For the sake of this case study, we narrow the scope to video on demand (VoD) monitored data (i.e., \Glb{Monitor} tools generating \textit{InfoMonitor} events) and textual feedback from social networks such as Twitter (i.e., \Glb{FeedbackGathering} tools generating \Glb{UserFeedback} events). This scenario is conceptualized in the UML depicted in Figure \ref{fig:UML}, which we use as a starting point to provide a high-level representation of the domain of interest that is later used to generate the ontological knowledge captured in $\mathcal{G}$. Figure \ref{fig:runningExampleGlobalOnt} in Section \ref{sec:ontology} depicts the RDF-based representation of the UML diagram used in our approach, which we will introduce in detail in that section.

Next, let us assume three data sources, in the form of REST APIs, and respectively one wrapper querying each. The first data source provides information related to the VoD monitor, which consist of JSON documents as depicted in Code \ref{code:JSONs_monitors}. We additionally define a wrapper on top of it obtaining the \Src{monitorId} of the monitor and computing the lag ratio metric (a quality of service measure computed as the fraction of wait and watch time) indicating the percentage of time a user is waiting for a video. The query of this wrapper is depicted in Code \ref{code:Q_mongodb} using MongoDB syntax\footnote{Note that the use of the \texttt{aggregate} keyword is used to invoke the aggregate querying framework. The \texttt{aggregate} keyword does not entail grouping unless the \texttt{\$group} keyword is used. Thus, note no aggregation is performed in this query.}, where for each tuple the attribute \Src{VoDmonitorId} (renamed from \Src{monitorId} in the JSON) and \Src{lagRatio} are projected (respectively mapping to the conceptual attributes \Glb{toolId} and \Glb{lagRatio}).

\begin{center}
\begin{minipage}{.35\textwidth}
  \begin{lstlisting}[language=json, captionpos=b, label={code:JSONs_monitors}, caption=Sample JSON for VoD monitors]
{
  "monitorId": 12,
  "timestamp": 1475010424,
  "bitrate": 6,
  "waitTime": 3,
  "watchTime": 4
}
\end{lstlisting}
\end{minipage}
\hfill
\begin{minipage}{.63\textwidth}
  \begin{lstlisting}[
    basicstyle=\footnotesize,language=json, label={code:Q_mongodb}, captionpos=b, caption=Wrapper projecting attributes \Src{VoDmonitorId} and \Src{lagRatio} (using MongoDB's Aggregation Framework syntax)]
db.getCollection("vod").aggregate([
  {$project: {
    "VoDmonitorId":"$monitorId", 
    "lagRatio": {$divide : ["$waitTime","$watchTime"]}}
  }  
])
\end{lstlisting}
\end{minipage}
\end{center}

For the sake of simplicity, hereinafter, we will represent wrappers as relations where their schema are the attributes projected by the queries, dismissing the details of the underlying query. Hence, the previous wrapper would be depicted as $w_1(\Src{VoDmonitorId},\Src{lagRatio})$ (note that the JSON key \Src{monitorId} has been renamed to \Src{VoDmonitorId}). To complete our running example, we define a wrapper $w_2(\Src{FGId}, \Src{tweet})$ providing, respectively, the \Glb{toolId} for the \Glb{FeedbackGathering} at hand and the \Glb{description} for such \Glb{UserFeedback}. Finally, the wrapper $w_3(\Src{TargetApp}, \Src{MonitorId}, \Src{FeedbackId})$ states for each \Glb{SoftwareApplication} the \Glb{toolId} of its associated \Glb{Monitor} and \Glb{FeedbackGathering} tools. Table \ref{tab:wrapper_samples} depicts an example of the output generated by each wrapper.

\begin{table}[h]
	\begin{minipage}{.35\textwidth}
	\footnotesize
    \begin{tabular}[h]{|c|c|}
    	\hline
    	\multicolumn{2}{|c|}{$w_1$}\\ \hline
    	\Src{VoDmonitorId}&\Src{lagRatio}\\ \hline
    	12&0.75\\ \hline
		12&0.90\\ \hline
		18&0.1\\ \hline
	\end{tabular}
    \end{minipage}
    \begin{minipage}{\textwidth}
	\begin{tabular}[h]{|c|c|}
    	\hline
    	\multicolumn{2}{|c|}{$w_2$}\\ \hline
    	\Src{FGId}&\Src{tweet}\\ \hline
    	77&``I continuously see the loading symbol''\\ \hline
		45&``Your video player is great!''\\ \hline
	\end{tabular}
	\end{minipage}
	\\
	\begin{minipage}{\textwidth}
	\centering
	\begin{tabular}[h]{|c|c|c|}
    	\hline
    	\multicolumn{3}{|c|}{$w_3$}\\ \hline
    	\Src{TargetApp}&\Src{MonitorId}&\Src{FeedbackId}\\ \hline
    	1&12&77\\ \hline
		2&18&45\\ \hline
	\end{tabular}
	\end{minipage}
	\caption{\label{tab:wrapper_samples}Sample output for each of the exemplary wrappers.}
\end{table}

Now, the goal is to enable data analysts to query the attributes of the ontology-based representation of the UML diagram (i.e., $\mathcal{G}$) by navigating over the classes, such that the sources are automatically accessed. Throughout the paper we will make use of the exemplary query retrieving for each \Glb{applicationId} its \Glb{lagRatio} instances. Hence, the task consists of rewriting such OMQ to an equivalent one over the wrappers, which can be translated to the following relational algebra expression: $\Pi_{w_3.\Src{TargetApp}, w_1.\Src{lagRatio}} (w_1 \underset{\Src{VoDmonitorId} = \Src{MonitorId}}{\bowtie} w_3)$. Table \ref{tab:query1} depicts an example of the output generated by such query.

\begin{table}[h]
	\centering
	\footnotesize
    \begin{tabular}[h]{|c|c|}
    	\hline
    	\Src{TargetApp}&\Src{lagRatio}\\ \hline
    	1&0.75\\ \hline
		1&0.90\\ \hline
		2&0.1\\ \hline
	\end{tabular}
	\caption{\label{tab:query1}Sample output for the exemplary query.}
\end{table}

Assume now that the first data source releases a new version of its API and in the new schema \Src{lagRatio} has been renamed to \Src{bufferingRatio}. Hence, a new wrapper $w_4(\Src{VoDmonitorId}, \Src{bufferingRatio})$ is defined. With such setting, the analyst should not be aware of such schema evolution, but now the query should consider both versions and be automatically rewritten to the following expression: $\Pi_{w_3.\Src{TargetApp}, w_1.\Src{lagRatio}} (w_1 \underset{\Src{VoDmonitorId} = \Src{MonitorId}}{\bowtie} w_3)$ $\bigcup$ $	\Pi_{w_3.\Src{TargetApp}, w_4.\Src{bufferingRatio}} (w_4 \underset{\Src{VoDmonitorId} = \Src{MonitorId}}{\bowtie} w_3)$.

\subsection{Notation}\label{sec:notation}

We consider a set of data sources $D = \{ D_1, \ldots, D_n \}$, where each $D_i$ consists of a set of wrappers $\{ w_1, \ldots, w_m \}$ representing views over different schema versions. We define the operator $source(w)$, which returns the data source $D$ to which $w$ belongs to. As previously stated, a wrapper is represented as a relation with the attributes its query projects. We distinguish between ID and non-ID attributes, hence a wrapper is defined as $w(\overline{a_{ID}}, \overline{a_{nID}})$, where $\overline{a_{ID}}$ and $\overline{a_{nID}}$ are respectively the set of its ID attributes and non-ID attributes.

\paragraph{Example} The VoD monitoring API would be depicted as $D_1 = \{ w_1(\{\Src{VoDmonitorId}\}, \{\Src{lagRatio}\}), w_4(\{\Src{VoDmonitorId}\}, \{\Src{bufferingRatio}\}) \}$, the feedback gathering API as $D_2 = \{ w_2(\{\Src{FGId}\}, \{\Src{tweet}\})$ and the relationship API as $D_3 = \{ w_3(\{\Src{TargetApp}, \Src{MonitorId}, \Src{FeedbackId}\}, \{\})$.
\\

Wrappers can be joined to each other by means of a restricted equi-join on IDs ($\widetilde{\bowtie}$). The semantics of $\widetilde{\bowtie}$ are those of an equi-join ($w_i \underset{a = b}{\bowtie} w_j$), but only valid if $a \in w_i.\overline{a_{ID}}$ and $b \in w_j.\overline{a_{ID}}$.  We also define the projection operator $\widetilde{\Pi}$, whose semantics are likewise a standard projection for non-ID attributes. We do not permit to project out any ID attribute, as they are necessary for $\widetilde{\bowtie}$. With such constructs, we can now define the concept of a walk over the wrappers ($W$), which consists of a relational algebra expression where wrappers are joined ($\widetilde{\bowtie}$) and their attributes are projected ($\widetilde{\Pi}$). Thus, we formally define a walk as $W = \widetilde{\Pi}(w_1) \widetilde{\bowtie} \ldots \widetilde{\bowtie} \widetilde{\Pi}(w_k)$. Furthermore, we work under the assumption that schema versions from the same data source should not be joined (e.g., $w_1$ and $w_4$ in the running example). To formalize this assumption let $wrappers(W)$ denote the set of wrappers used in walk $W$. Then we require that $\forall w_i, w_j \in wrappers(W) : source(w_i) \neq source(w_j)$. Note that a walk can also be seen as a conjunctive query over the wrappers (i.e., select-project-join expression), thus two walks are equivalent if they join the same wrappers dismissing the order how this is done. Consider, however, that as the operator $\widetilde{\Pi}$ does not project out ID attributes, all ID attributes will be part of the output schema.

\paragraph{Example} The exemplary query (i.e., for each \Glb{applicationId} fetch its \Glb{lagRatio} instances) would consist of two walks $W_1 = \widetilde{\Pi}_{\Src{lagRatio}}(w_1)$ $\underset{\Src{VoDmonitorId} = \Src{MonitorId}}{\widetilde{\bowtie}}$ $ \widetilde{\Pi}_{\Src{TargetApp}}(w_3)$ and $W_2 = \widetilde{\Pi}_{\Src{bufferingRatio}}(w_4)$ $\underset{\Src{VoDmonitorId} = \Src{MonitorId}}{\widetilde{\bowtie}}$ $\widetilde{\Pi}_{\Src{TargetApp}}(w_3)$.
\\

Next, we formalize the ontology $\mathcal{T}$ as a 3-tuple $\langle \mathcal{G}, \mathcal{S}, \mathcal{M} \rangle$ of RDF named graphs. The Global graph ($\mathcal{G}$) contains the concepts and relationships that analysts will use to query, the source graph ($\mathcal{S}$) the data sources and the schemata of wrappers, and the mappings graph ($\mathcal{M}$) the LAV mappings between $\mathcal{S}$ and $\mathcal{G}$. Recall that data analysts pose OMQs over $\mathcal{G}$, however we do not allow arbitrary queries. We restrict OMQs to a subset of standard SPARQL defining subgraph patterns of $\mathcal{G}$, and only project elements of such pattern. Code \ref{code:Q_pattern} depicts the template of the permitted queries. Precisely, $attr_1, \ldots, attr_n$ must be attribute URIs (i.e., mapping to the UML attributes in Fig. \ref{fig:UML}), where each $attr_i$ has an invited variable $?v_i$ in the SELECT clause. The set of triples in the WHERE clause must define a connected subgraph of $\mathcal{G}$. On the one hand, it contains triples of the form $\langle s_i, hasFeature, attr_i \rangle$, where $s_i$ are class URIs (i.e., mapping to UML classes) and $hasFeature$ a predicate stating that $attr_i$ is attribute of class $s_i$. On the other hand, it contains triples of the form $\langle s_j, p_j, o_j \rangle$, where $s_j$ and $o_j$ are class URIs (i.e., mapping to UML classes) and $p_i$ predicate URIs (i.e., mapping to relationships between UML classes).

\begin{lstlisting}[basicstyle=\small, language=SQL, captionpos=b, label={code:Q_pattern}, caption=Template for accepted SPARQL queries,mathescape=true]
SELECT $?v_1$ $\ldots$ $?v_n$
FROM $\mathcal{G}$
WHERE {
 VALUES ($?v_1$ $\ldots$ $?v_n$) { ($attr_1$ $\ldots$ $attr_n$) }
 $s_1$ $p_1$ $attr_1$ .
 $\ldots$
 $s_n$ $p_n$ $attr_n$ .
 $\ldots$
 $s_m$ $p_m$ $o_m$
}
\end{lstlisting}

OMQs are meant to be translated to sets of walks, to this end the aforementioned SPARQL queries must be parsed and manipulated. This task can be simplified leveraging on SPARQL Algebra\footnote{\url{https://www.w3.org/2001/sw/DataAccess/rq23/rq24-algebra.html}}, where the semantics of the query evaluation are specified. Libraries such as ARQ\footnote{\url{https://www.w3.org/2011/09/SparqlAlgebra/ARQalgebra}} provide mechanisms to get such algebraic structure for a given SPARQL query. Code \ref{code:ARQ} depicts the algebra structure generated after parsing the subset of permitted SPARQL queries.

\begin{lstlisting}[basicstyle=\small, captionpos=b, label={code:ARQ}, caption=SPARQL algebra for the accepted SPARQL queries,mathescape=true]
(project ($?v_1$ $\ldots$ $?v_n$)
  (join
    (table (vars $?v_1$ $\ldots$ $?v_n$)
      (row [$?v_1$ $attr_1$] $\ldots$ [$?v_n$ $attr_n$])
    )
    (bgp
      (triple $s_1$ $p_1$ $attr_1$)
      $\ldots$
      (triple $s_n$ $p_n$ $attr_n$)
      $\ldots$
      (triple $s_m$ $p_m$ $o_m$)
)))))
\end{lstlisting}

In order to easily manipulate such algebraic structures, we formalize the allowed SPARQL queries as $Q_{\mathcal{G}} = \langle \pi, \varphi \rangle$, where $\pi$ is the set of projected attributes (i.e., the URIs $attr_1$, $\ldots$, $attr_n$) and $\varphi$ the graph pattern specified under the \texttt{bgp} clause (i.e., basic graph pattern). Note that $\pi \subseteq V(\varphi)$, where $V(\varphi)$ returns the vertex set of $\varphi$. 

\paragraph{Example} The exemplary query is depicted using SPARQL in Code \ref{code:Q_example}. Alternatively, it would be represented as $\pi = \{ lagRatio, applicationId \}$, and $\varphi$ the subgraph $applicationId \xleftarrow[hasFeature]{} SoftwareApplication \xrightarrow[hasMonitor]{} Monitor \xrightarrow[generatesQoS]{} InfoMonitor \xrightarrow[hasFeature]{} lagRatio$.

\begin{lstlisting}[basicstyle=\small, language=SQL, captionpos=b, label={code:Q_example}, caption=Running example's SPARQL query,mathescape=true]
SELECT $?x$ $?y$
FROM $\mathcal{G}$
WHERE {
 VALUES ($?x$ $?y$) { ($applicationId$ $lagRatio$) }
 $SoftwareApplication$ $hasFeature$ $applicationId$ $.$
 $SoftwareApplication$ $hasMonitor$ $Monitor$ $.$
 $Monitor$ $generatesQoS$ $InfoMonitor$ $.$
 $InfoMonitor$ $hasFeature$ $lagRatio$
}
\end{lstlisting}

The wrappers and the ontology are linked by means of schema mappings. Those are commonly formalized using tuple-generating dependencies (tgds) \cite{DBLP:journals/tcs/FaginKMP05}, which are logical expressions of the form $\forall x (\exists y \Phi(x,y) \mapsto \exists z \Psi(x,z))$, where $\Phi$ and $\Psi$ are conjunctive queries. However, in our context we serialize such mappings in the graph $\mathcal{M}$, and not as separated logical expressions. Hence, we define a LAV mapping for a wrapper $w$ as $LAV(w): w \mapsto \varphi_{\mathcal{G}}$, where $\varphi_{\mathcal{G}}$ is a subgraph of $\mathcal{G}$. We additionally consider a function $F: a_w \mapsto a_m$, that translates the name of an attribute in $\mathcal{S}$ to its corresponding conceptual representation in $\mathcal{G}$. Such function allows us to denote semantic equivalence between physical and conceptual attributes in the ontology (respectively, in $\mathcal{S}$ and $\mathcal{G}$). Intuitively, $F$ forces a physical attribute in the sources to map to one and only one conceptual feature in $\mathcal{G}$. As schema mappings, this function is also serialized in $\mathcal{M}$.

\paragraph{Example} The LAV mapping for $w_1$ would be the subgraph $Monitor \xrightarrow[generatesQoS]{} InfoMonitor$ (also including all class attributes). Regarding $F$, the function would make the conversions $w_1.\Src{VoDmonitorId} \mapsto toolId$ and $w_1.\Src{lagRatio} \mapsto lagRatio$.

\subsection{Problem statement}

In order to introduce the problem statement we must first introduce the notions of \textit{coverage} and \textit{minimality} for a query $Q_{\mathcal{G}}$ over $\mathcal{G}$ and a walk $W$. \textit{Coverage} is formalized as $\bigcup_{w \in wrappers(W)} LAV(w) \supseteq Q_{\mathcal{G}}$, which states that a walk covers the query if the union of the LAV graphs of the wrappers participating in the walk subsume $Q_{\mathcal{G}}$. \textit{Minimality} is formalized as $\forall_{w \in W} (coverage(W,Q_{\mathcal{G}}) \wedge \neg coverage(W \setminus w,Q_{\mathcal{G}}) )$, which states that if any wrapper is removed from a covering walk, then the walk is not covering anymore. Intuitively, these properties guarantee that a walk answering a query contains all the required attributes and joins, and each wrapper contributes with at least one attribute.

Now, with the previously introduced formalization and properties, we can state the problem of ontology-based query answering under LAV mappings as a faceted search over the wrappers with the goal of finding all possible ways to obtain the requested attributes. Given an OMQ $Q_{\mathcal{G}}$, we aim at finding a set of non-equivalent walks $\mathcal{W}$ such that each $W \in \mathcal{W}$ is \textit{covering} and \textit{minimal} with respect to $Q_{\mathcal{G}}.\varphi$. As a result, we obtain a union of conjunctive queries, which corresponds to the union of all the covering and minimal walks found for $Q_{\mathcal{G}}.\varphi$.
\section{Big Data Integration ontology} \label{sec:ontology}

In this section, we present the Big Data Integration ontology (BDI), the metadata artifact that enables a systematic approach for the data integration system governance when ingesting and analysing the data. To this end, we have followed the well-known theory on data integration \cite{DBLP:conf/pods/Lenzerini02} and divided it into two levels (by means of RDF named graphs): the Global and Source graphs, respectively $\mathcal{G}$ and $\mathcal{S}$, linked via mappings $\mathcal{M}$. Thanks to the extensibility of RDF, it further enables us to enrich $\mathcal{G}$ and $\mathcal{S}$ with semantics such as data types. In this section we present the RDF vocabulary to be used to represent $\mathcal{G}$ and $\mathcal{S}$. To do so, we present a metamodel for the global and source ontologies that current models (i.e., $\mathcal{G}$ and $\mathcal{S}$) must mandatorily follow. In the following subsections, we elaborate on each graph and present its RDF representation.

\subsection{Global graph}

The Global graph $\mathcal{G}$ reflects the main domain concepts, relationships among them and features of analysis (i.e., maps to the role of a UML diagram in a machine-readable format). Its elements are defined in terms of the vocabulary users will use when posing queries. The metadata model for $\mathcal{G}$ distinguishes concepts from features, the former mimicking classes and the latter attributes in a UML diagram. Concepts can be linked by means of domain-specific object properties, which implicitely determine their domain and range. Such properties will be used for data analysts to navigate the graph, dismissing the need of specifying how the underlying sources are joined.
The link between a concept and its set of features is represented via \Rdf{G:hasFeature}. In order to disambiguate the query rewriting process we restrict features to belong to only one concept. Additionally, it is possible to define a taxonomy of features, which will denote related semantic domains (e.g., the feature \Rdf{sup:monitorId} is subclass of \Rdf{sc:identifier}). Features can be enriched with new semantics to aid the data management and analysis phases. In this paper, we narrow the scope to data types for features, widely used in data integrity management.

Code \ref{code:metamodelG} provides the triples that compose $\mathcal{G}$ in Turtle RDF notation\footnote{\url{https://www.w3.org/TR/turtle}}. It contains the main metaclasses (using the namespace prefix \Rdf{G}\footnote{\url{http://www.essi.upc.edu/~snadal/BDIOntology/Global}} as main namespace) which all features of analysis will instantiate. Concepts and features can reuse existing vocabularies by following the principles of the Linked Data (LD) initiative.
Additionally, we include elements for data types on features linked using \Rdf{G:hasDatatype}, albeit their maintenance is out of the scope of this paper. Following the same LD philosophy, we reuse the \Rdf{rdfs:Datatype} class to instantiate data types. With such design, we favor the elements of $\mathcal{G}$ to be of any of the available types in XML Schema (prefix \Rdf{xsd}\footnote{\url{http://www.w3.org/2001/XMLSchema}}). Finally, note that we focus on non-complex data types, however our model can be easily extended to include complex types \cite{1_downey_2005}.
\\

\begin{lstlisting}[language=turtle, captionpos=b, label={code:metamodelG}, caption=Metadata model for $\mathcal{G}$ in Turtle notation]
@prefix owl: <http://www.w3.org/2002/07/owl#> .
@prefix rdf: <http://www.w3.org/1999/02/22-rdf-syntax-ns#> .
@prefix rdfs: <http://www.w3.org/2000/01/rdf-schema#> .
@prefix voaf: <http://purl.org/vocommons/voaf#> .
@prefix vann: <http://purl.org/vocab/vann/> .
@prefix G: <http://www.essi.upc.edu/~snadal/BDIOntology/Global/> . 

<http://www.essi.upc.edu/~snadal/BDIOntology/Global/> rdf:type voaf:Vocabulary ;
        vann:preferredNamespacePrefix "G";
        vann:preferredNamespaceUri "http://www.essi.upc.edu/~snadal/BDIOntology/Global";
        rdfs:label "The Global graph vocabulary" .

G:Concept rdf:type rdfs:Class;
        rdfs:isDefinedBy <http://www.essi.upc.edu/~snadal/BDIOntology/Global/> .

G:Feature rdf:type rdfs:Class;
        rdfs:isDefinedBy <http://www.essi.upc.edu/~snadal/BDIOntology/Global/> .

G:hasFeature rdf:type rdf:Property ;
        rdfs:isDefinedBy <http://www.essi.upc.edu/~snadal/BDIOntology/Global/> ;
        rdfs:domain G:Concept ;
        rdfs:range G:Feature .
        
G:hasDataType rdf:type rdf:Property ;
        rdfs:isDefinedBy <http://www.essi.upc.edu/~snadal/BDIOntology/Global/> ;
        rdfs:domain G:Feature ;
        rdfs:range rdfs:Datatype .
\end{lstlisting}

\begin{figure*}
\centering
\includegraphics[width=1\linewidth]{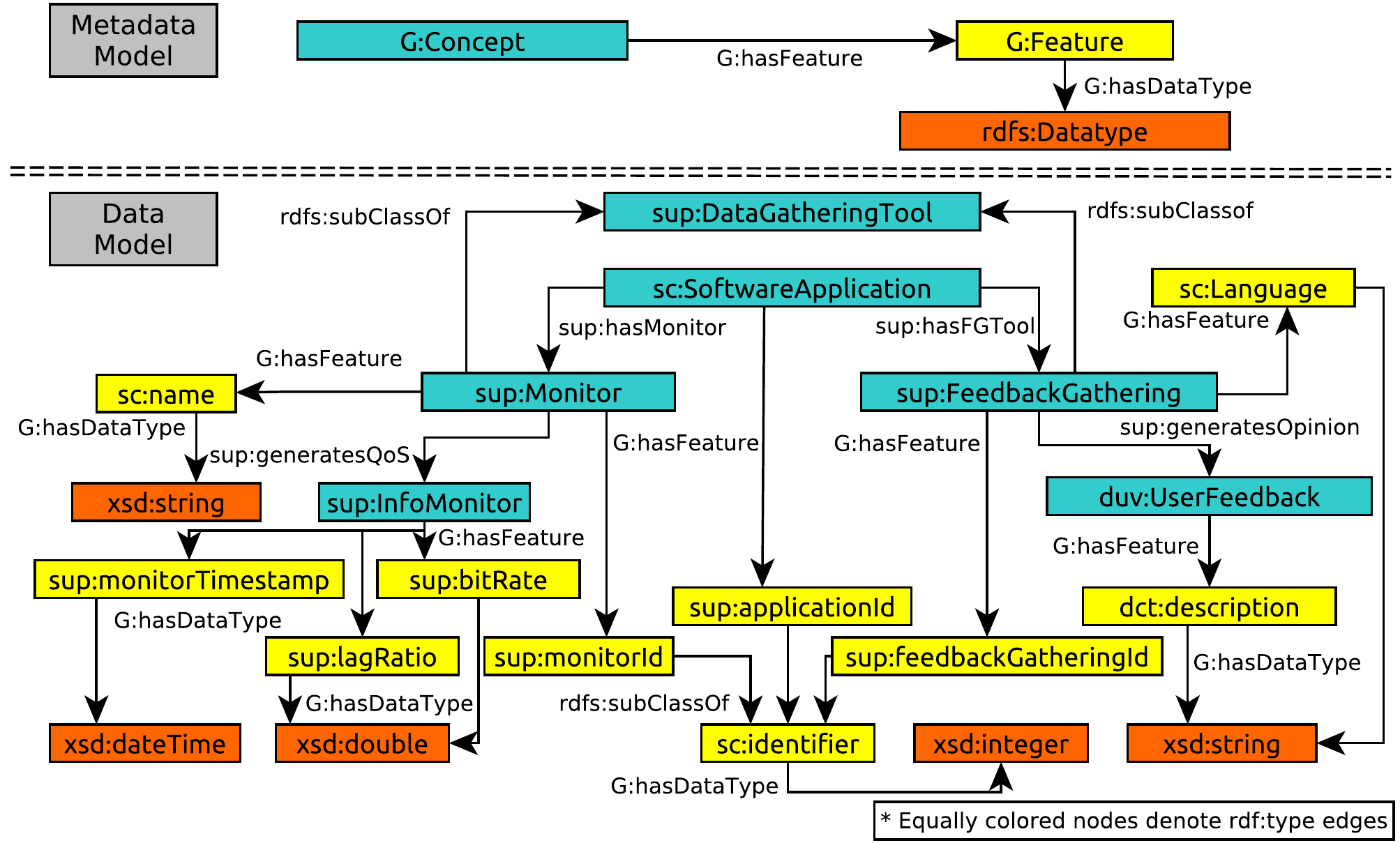}
\caption{RDF dataset of the metadata model and data model of $\mathcal{G}$ for the SUPERSEDE running example. For interpretation of the references to color in the text, the reader is referred to the web version of this article.} \label{fig:runningExampleGlobalOnt}
\end{figure*}

\paragraph{Example} Figure \ref{fig:runningExampleGlobalOnt} depicts the instantiation of $\mathcal{G}$ in the SUPERSEDE case study, as presented in the UML diagram in Figure \ref{fig:UML} (for the sake of conciseness only a fragment is depicted). The color of the elements represent typing (i.e., \Rdf{rdf:type} links). Note that, in order to comply with the design constraints of $\mathcal{G}$ (i.e., a feature can only belong to one concept), the \Glb{toolId} feature has been explicited and made distinguishable to \Rdf{sup:monitorId} and \Rdf{sup:feedbackGatheringId} respectively for classes \Glb{Monitor} and \Glb{FeedbackGathering}. When possible, vocabularies are reused, namely \url{https://www.w3.org/TR/vocab-duv} (prefix \Rdf{duv}) for feedback elements as well as \url{http://dublincore.org/documents/dcmi-terms} (prefix \Rdf{dct}) or \url{http://schema.org} (prefix \Rdf{sc}). However, when no vocabulary is available we define the custom SUPERSEDE vocabulary (prefix \Rdf{sup}).

\subsection{Source graph}

The purpose of the Source graph $\mathcal{S}$ is to model the different wrappers and their provided schema. To this end, we define the metaconcept \Rdf{S:DataSource} which models the different data sources (e.g., Twitter REST API). In $\mathcal{S}$, we additionally encode the necessary information for schema versioning, hence we define the metaconcept \Rdf{S:Wrapper} which will model the different schema versions for a data source, which in turn consist of a representation of the projected attributes, modeled in the metaconcept \Rdf{S:Attribute}. We embrace the reuse of attributes within wrappers of the same data source, as we assume the semantics do not differ across schema versions, however that assumption is not realistic among different data sources (e.g., not necessarily a timestamp has the same meaning in the VoD monitor and the Twitter API). Therefore, we encode in the attribute names the prefix of the data source they correspond to (e.g., for a data source $D$, its wrappers $W$ and $W^{\prime}$ respectively provide the attributes $\{ \Rdf{D/a}, \Rdf{D/b} \}$ and $\{ \Rdf{D/a}, \Rdf{D/c} \}$). Code \ref{code:metamodelS} depicts the metadata model for $\mathcal{S}$ in Turtle RDF notation (using prefix \Rdf{S}\footnote{\url{http://www.essi.upc.edu/~snadal/BDIOntology/Source}} as main namespace).
\\

\begin{lstlisting}[language=turtle, captionpos=b, label={code:metamodelS}, caption=Metadata model for $\mathcal{S}$ in Turtle notation]
@prefix owl: <http://www.w3.org/2002/07/owl#> .
@prefix rdf: <http://www.w3.org/1999/02/22-rdf-syntax-ns#> .
@prefix rdfs: <http://www.w3.org/2000/01/rdf-schema#> .
@prefix voaf: <http://purl.org/vocommons/voaf#> .
@prefix vann: <http://purl.org/vocab/vann/> .
@prefix S: <http://www.essi.upc.edu/~snadal/BDIOntology/Source/> . 

<http://www.essi.upc.edu/~snadal/BDIOntology/Source/> rdf:type voaf:Vocabulary ;
        vann:preferredNamespacePrefix "S";
        vann:preferredNamespaceUri "http://www.essi.upc.edu/~snadal/BDIOntology/Source";
        rdfs:label "The Source graph vocabulary" .

S:DataSource rdf:type rdfs:Class;
        rdfs:isDefinedBy <http://www.essi.upc.edu/~snadal/BDIOntology/Source/> .

S:Wrapper rdf:type rdfs:Class;
        rdfs:isDefinedBy <http://www.essi.upc.edu/~snadal/BDIOntology/Source/> .

S:Attribute rdf:type rdfs:Class;
        rdfs:isDefinedBy <http://www.essi.upc.edu/~snadal/BDIOntology/Source/> .

S:hasWrapper rdf:type rdf:Property ;
        rdfs:isDefinedBy <http://www.essi.upc.edu/~snadal/BDIOntology/Source/> ;
        rdfs:domain S:DataSource ;
        rdfs:range S:Wrapper .		

S:hasAttribute rdf:type rdf:Property ;
        rdfs:isDefinedBy <http://www.essi.upc.edu/~snadal/BDIOntology/Source/> ;
        rdfs:domain S:Wrapper ;
        rdfs:range S:Attribute .
\end{lstlisting}

\begin{figure*}[h!]
\centering
\includegraphics[width=1\linewidth]{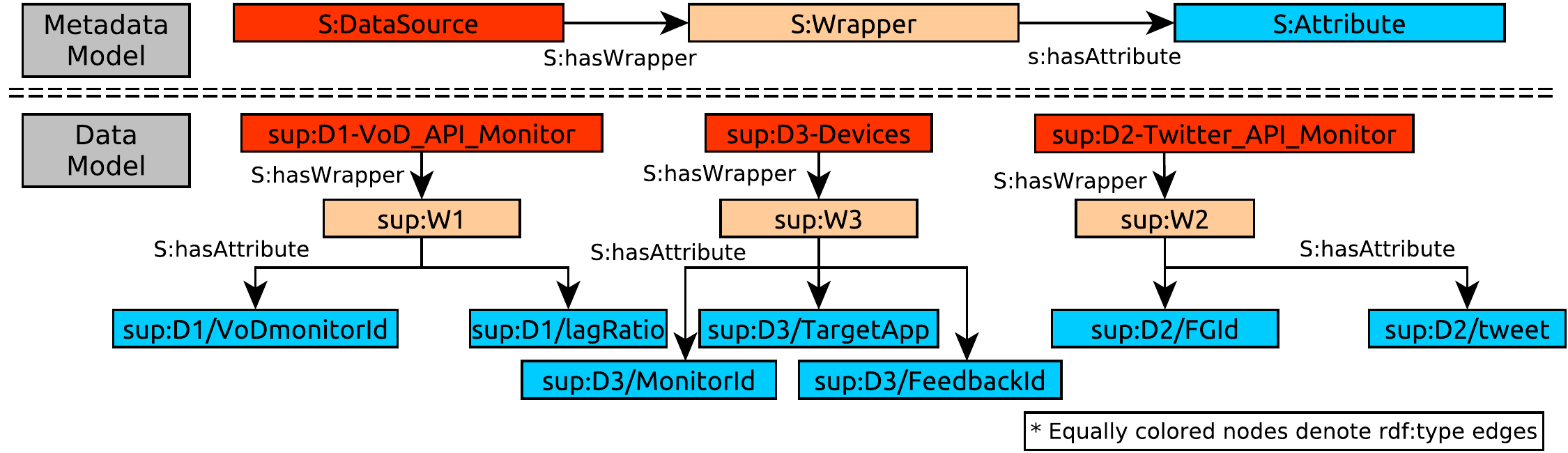}
\caption{RDF dataset of the metadata model and data model of $\mathcal{S}$. For interpretation of the references to color in the text, the reader is referred to the web version of this article.} \label{fig:runningExampleSourceOnt}
\end{figure*}

\paragraph{Example} Figure \ref{fig:runningExampleSourceOnt} shows the instantiation of $\mathcal{S}$ in SUPERSEDE. Red nodes depict the data sources that correspond to the three data sources introduced in Section \ref{sec:runningExample}. Then, orange and blue nodes depict the wrappers and attributes, respectively.

\subsection{Mapping graph}\label{sec:mappings_graph}

As previously discussed, we encode LAV mappings in the ontology. Recall that mappings are composed by (a) subgraphs of $\mathcal{G}$, one per wrapper, and (b) the function $F$ linking elements of type \Rdf{S:Attribute} to elements of type \Rdf{G:Feature}. We serialize such information in RDF in the Mapping graph $\mathcal{M}$. Subgraphs are represented using named graphs, which identify a subset of $\mathcal{G}$. Thus, each wrapper will have associated a named graph identifying which concepts and features it is providing information about. This will be represented using triples of the form $\langle w, \Rdf{M:mapping}, G \rangle$, where $w$ is an instance of \Rdf{S:Wrapper} and G is a subgraph of $\mathcal{G}$. Regarding the function $F$, we represent it via the \Rdf{owl:sameAs} property (i.e., triples of the form $\langle \Src{x}, \Rdf{owl:sameAs}, \Glb{y} \rangle$, where $\Src{x}$ and $\Glb{y}$ are respectively instances of \Rdf{S:Attribute} and \Rdf{G:Feature}.

\paragraph{Example} 

\begin{figure*}[h!]
\centering
\includegraphics[width=1\linewidth]{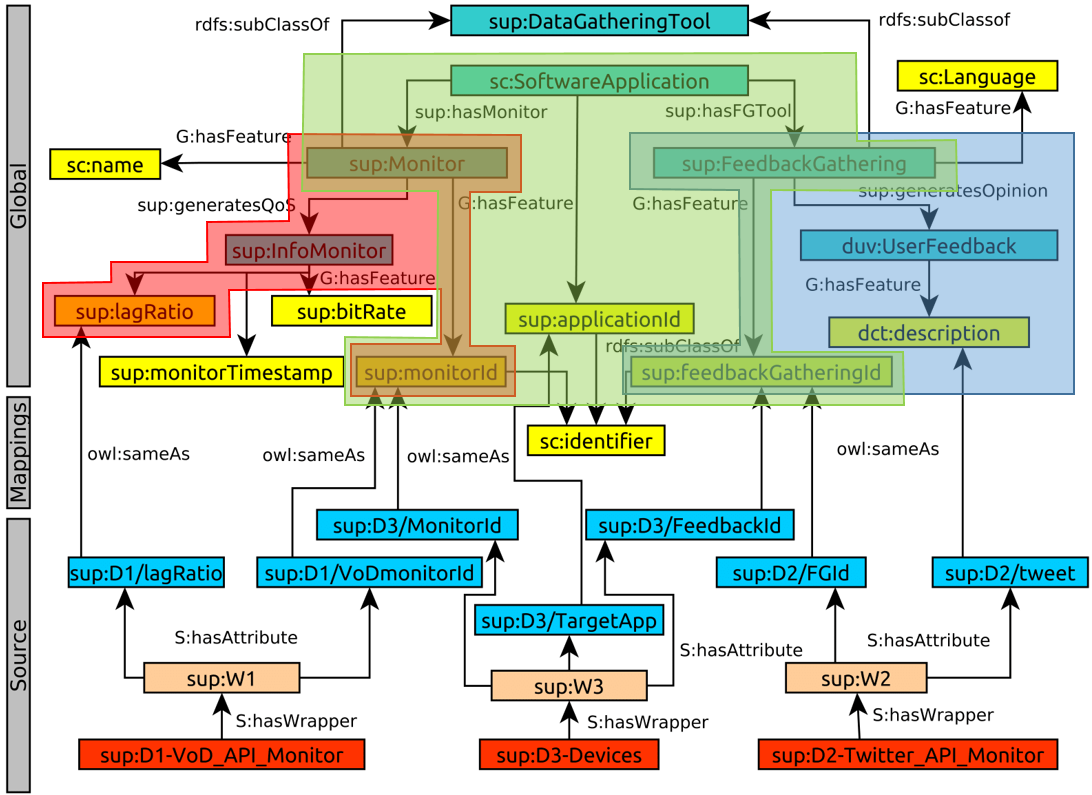}
\caption{RDF dataset of the metadata model and data model of the complete ontology for the SUPERSEDE running example. For interpretation of the references to color in the text, the reader is referred to the web version of this article.} \label{fig:runningExampleMappings}
\end{figure*}

In Figure \ref{fig:runningExampleMappings} we depict the complete instantiation of the BDI ontology for the SUPERSEDE running example. To ensure readability, internal classes are omitted and only the core ones are shown. Named graphs are depicted using colored boxes, respectively red for $w_1$, blue for $w_2$ and green for $w_3$.

The previous discussion sets the baseline to enable semi-automatic schema management in the data sources. Instantiating the metadata model, the data steward is capable of modeling the schema of the sources to be further linked to the wrappers and the data instances they provide. With such, in the rest of this paper we will introduce techniques to adapt the ontology to schema evolution aswell as query answering.
\section{Handling evolution} \label{sec:evolution}

In this section, we present how the BDI ontology accomodates the evolution of situational data. Specific studies concerning REST API evolution \cite{DBLP:conf/icws/LiXLZ13,DBLP:conf/icsoc/WangKZ14} have concluded that most of such changes occur in the structure of incoming events, thus our goal is to semi-automatically adapt the BDI ontology to such evolution. To this end, in the following subsections we present an algorithm to aid the data steward to enrich the ontology upon new releases.

\subsection{Releases}

In Section \ref{sec:overview}, we discussed the role of the data steward as the unique maintainer of the BDI ontology in order to make data management tasks transparent to data analysts. 
Now, the goal is to shield the analysts queries, so that they do not crash upon new API version releases. In other words, we need to adapt $\mathcal{S}$ to schema evolution in the data sources, so that $\mathcal{G}$ is not affected. To this end, we introduce the notion of \textit{release}, the construct indicating the creation of a new wrapper, and how its elements link to features in $\mathcal{G}$. Thus, we formally define a release $R$ as a 3-tuple $R = \langle w, G, F \rangle$, where $w$ is a wrapper, $G$ is a subgraph of  $\mathcal{G}$ denoting the elements in $\mathcal{G}$ that the wrapper contributes to, and $F = a \mapsto V(G)$ a function where $a \in w.\overline{a_{ID}} \cup w.\overline{a_{nID}}$ and $V(G)$ vertices of type $\Rdf{G:Feature}$ in $\mathcal{G}$. $R$ must be created by the data steward upon new releases. Several approaches can aid this process. For instance, to define the graph $G$, the user can be presented with subgraphs of $\mathcal{G}$ that cover all features. However, this raises the question of which is the most appropiate subgraph that the user is interested in. Regarding the definition of $F$, probabilistic methods to align and match RDF ontologies, such as \textsc{paris} \cite{DBLP:journals/pvldb/SuchanekAS11}, can be used. Note that the definition of wrappers (i.e., how to query an API) is beyond the scope of this paper.

\paragraph{Example} Recall wrapper $w_4$ for data source $D_1$. Its associated release would be defined as $w_4(\Src{VoDmonitorId}, \Src{bufferingRatio})$, $G = \Rdf{sup:lagRatio} \xleftarrow[\Rdf{G:hasFeature}]{} \Rdf{sup:InfoMonitor} \xrightarrow[\Rdf{sup:hasMonitor}]{} \Rdf{sup:Monitor} \xrightarrow[\Rdf{G:hasFeature}]{} \Rdf{sup:monitorId}$, and $F = \{ \Src{VoDmonitorId} \mapsto \Rdf{sup:monitorId}, \Src{bufferingRatio} \mapsto \Rdf{sup:lagRatio}\}$.

\subsection{Release-based Ontology Evolution} \label{sec:algorithmEvolution}

As mentioned above, changes in the source elements need to be reflected in the ontology to avoid queries to crash. Furthermore, the ultimate goal is to provide such adaptation in an automated way. To this end, Algorithm \ref{algo:newRelease} applies the necessary changes to adapt the BDI ontology $\mathcal{T}$ w.r.t. a new release $R$. It starts registering the data source, in case it is new (line 4), and the new wrapper to further link them (lines 7 and 8). Then, for each attribute in the wrapper $R.w$, we check their existence in the current Source graph and register it, in case it is not present. Given the way URIs for attributes are constructed (i.e., they have the prefix of their source), we can ensure that only attributes from the same source will be reused within subsequent versions. This helps to maintain a low growth rate for $\mathcal{T}.\mathcal{S}$, as well as avoiding potential semantic differences. Next, the named graph is registered to the Mapping graph, to conclude with the serialization of function $F$ (in $R.F$). The complexity of this algorithm is linearly bounded by the size of the parameters of $R$.

\begin{algorithm}[!h]\scriptsize
\caption{\label{algo:newRelease} Adapt to Release}
\begin{algorithmic}[1]
\Require $\mathcal{T}$ is the BDI ontology, $R$ new release
\Ensure $\mathcal{T}$ is adapted w.r.t. $R$
\Function{NewRelease}{$\mathcal{T}$, $R$}
	\State $Source_{uri} = \Rdf{"S:DataSource/"} \Plus source(R.w)$
	\If{$Source_{uri} \notin \textrm{SELECT } ?ds \textrm{ FROM } \mathcal{T} \textrm{ WHERE } \langle ?ds, \Rdf{"rdf:type"}, \Rdf{"S:DataSource"} \rangle$}
		\State $\mathcal{T}.\mathcal{S} \ue \langle Source_{uri}, \Rdf{"rdf:type"}, \Rdf{"S:DataSource"} \rangle$	
	\EndIf
	\State $Wrapper_{uri} = \Rdf{"S:Wrapper/"} \Plus R.w$
	\State $\mathcal{T}.\mathcal{S} \ue \langle Wrapper_{uri}, \Rdf{"rdf:type"}, \Rdf{"S:Wrapper"} \rangle$
	\State $\mathcal{T}.\mathcal{S} \ue \langle Source_{uri}, \Rdf{"S:hasWrapper"}, Wrapper_{uri} \rangle$
	\ForAll{$a \in (R.w.\overline{a_{ID}} \cup R.w.\overline{a_{nID}})$}
		\State $Attribute_{uri} = Source_{uri} \Plus a$
		\If{$Attribute_{uri} \notin \textrm{SELECT } ?a \textrm{ FROM } \mathcal{T} \textrm{ WHERE } \langle ?a, \Rdf{"rdf:type"}, \Rdf{"S:Attribute"} \rangle$}
			\State $\mathcal{T}.\mathcal{S} \ue \langle Attribute_{uri}, \Rdf{"rdf:type"}, \Rdf{"S:Attribute"} \rangle$
		\EndIf
		\State $\mathcal{T}.\mathcal{S} \ue \langle Wrapper_{uri}, \Rdf{"S:hasAttribute"}, Attribute_{uri} \rangle$
	\EndFor
	\State $\mathcal{T}.\mathcal{M} \ue \langle Wrapper_{uri}, \Rdf{"M:mapping"}, R.G \rangle$
	\ForAll{$(a,f) \in R.F$}
		\State $a_{uri} = Source_{uri} \Plus a$
		\State $f_{uri} = \Rdf{"G:Feature/"} \Plus f$
		\State $\mathcal{T}.\mathcal{M} \ue \langle a_{uri}, \Rdf{"owl:sameAs"}, f_{uri} \rangle$
	\EndFor
\EndFunction
\end{algorithmic}
\end{algorithm}

\paragraph{Example}

\begin{figure*}[!h]
\centering
\includegraphics[width=1\linewidth]{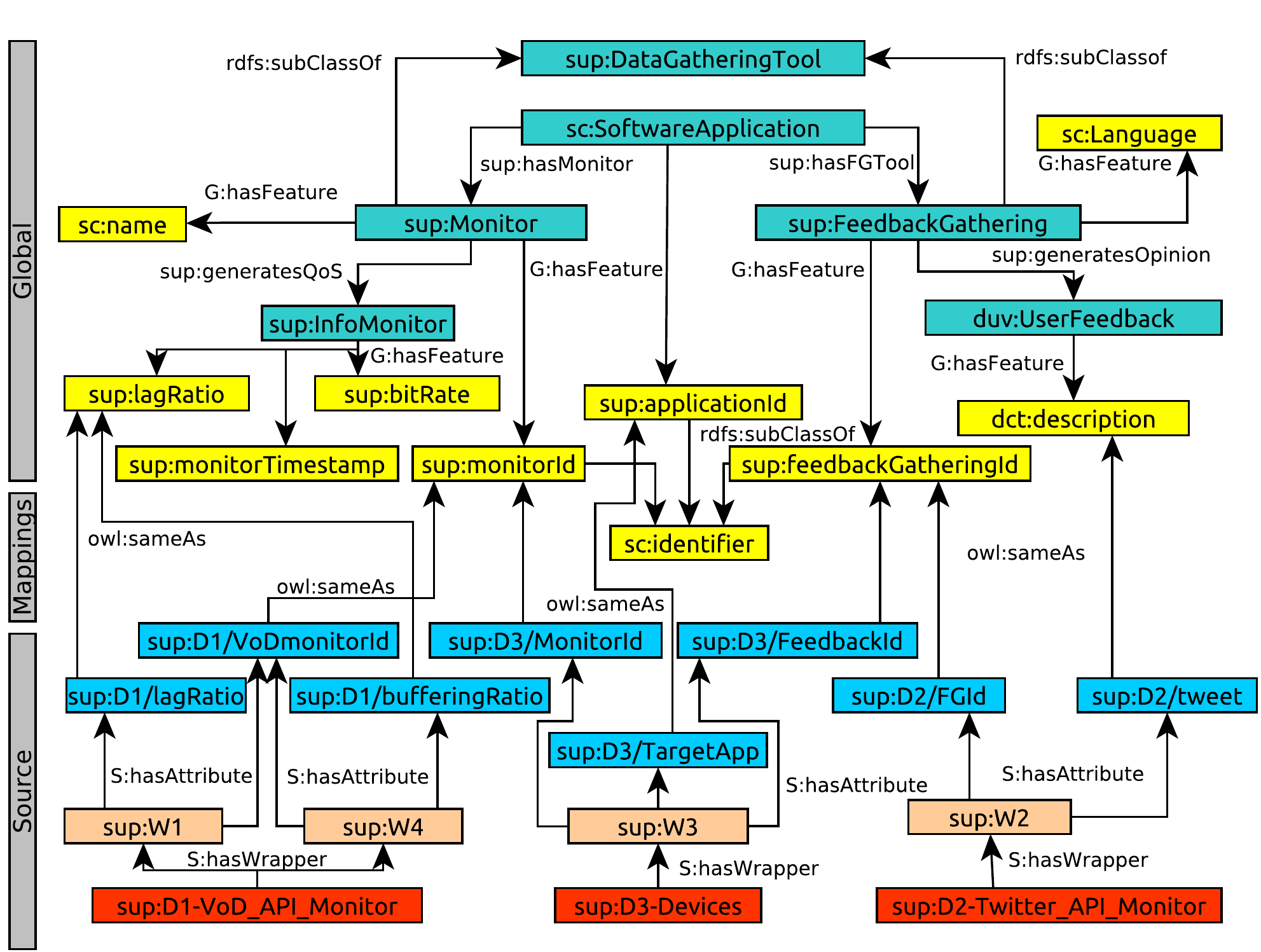}
\caption{RDF dataset for the evolved ontology $\mathcal{T}$ for the SUPERSEDE running example} \label{fig:evolution}
\end{figure*}

In Figure \ref{fig:evolution}, we depict the resulting ontology $\mathcal{T}$ after executing Algorithm \ref{algo:newRelease} with the release for wrapper $w_4$.

\section{Query answering} \label{sec:query_answering}

\begin{figure*}[h!]
\centering
\includegraphics[width=1\linewidth]{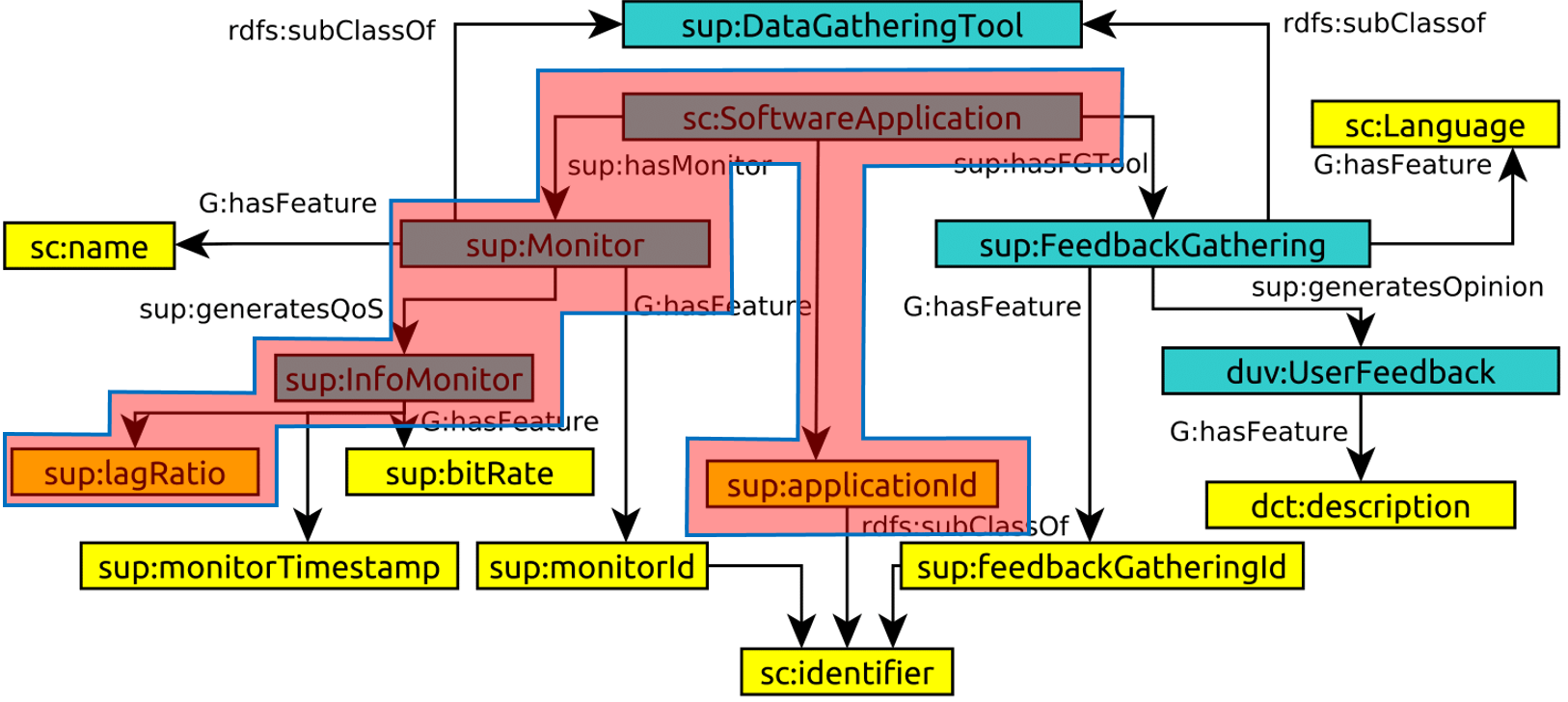}
\caption{Graph pattern for the running example query} \label{fig:queryGlobal}
\end{figure*}

In this section, we present the algorithm for ontology-based query answering under LAV mappings with wrappers. To this end, we provide a query rewriting algorithm that, given a conjunctive query $Q_{\mathcal{G}}$ produces a union of conjunctive queries $Q$ over the wrappers. Retaking the running example, and now using the vocabulary introduced in Section \ref{sec:ontology} as prefixes, the SPARQL representation of the query obtaining for each \Glb{applicationId} all its \Glb{lagRatio} instances would be that depicted in Code \ref{code:Q_query_example}. Alternatively, recall the alternative representation for $Q_{\mathcal{G}}$ as $Q_{\mathcal{G}}.\pi = \{\Rdf{sup:applicationId}, \Rdf{sup:lagRatio}\}$ and the graph $Q_{\mathcal{G}}.\varphi$ depicted in Figure \ref{fig:queryGlobal}.

\begin{lstlisting}[basicstyle=\small, language=SQL, captionpos=b, label={code:Q_query_example}, mathescape=true, caption=Running example's SPARQL query]
SELECT $?x$ $?y$
FROM $\mathcal{G}$
WHERE {
  VALUES ($?x$ $?y$) { ($\Rdf{sup:applicationId}$ $\Rdf{sup:lagRatio}$) }
  $\Rdf{sc:SoftwareApplication}$ $\Rdf{G:hasFeature}$ $\Rdf{sup:applicationId}$ $\Rdf{.}$
  $\Rdf{sc:SoftwareApplication}$ $\Rdf{sup:hasMonitor}$ $\Rdf{sup:Monitor}$ $\Rdf{.}$
  $\Rdf{sup:Monitor}$ $\Rdf{sup:generatesQoS}$ $\Rdf{sup:InfoMonitor}$ $\Rdf{.}$
  $\Rdf{sup:InfoMonitor}$ $\Rdf{G:hasFeature}$ $\Rdf{sup:lagRatio}$
}
\end{lstlisting}

\subsection{Well-formed queries}

As previously mentioned, unambiguously resolving query answering under LAV mappings entails constraining the design of the elements in the ontology, which also applies for the case of queries. Even though our approach makes transparent to the user how the concepts in $\mathcal{G}$ are to be joined in the wrappers, it is necessary that $Q.\pi$ retrieves only elements that exist in the sources (i.e., features) and can be populated with data. To this end, we introduce the notion of well-formed query.

\begin{definition}[Well-formed query]
A query $Q_{\mathcal{G}}$ is well formed iff $Q_{\mathcal{G}}.\varphi$ has a topological sorting (i.e., it is a DAG) and any projected element $p \in Q_{\mathcal{G}}.\pi$ refers to a terminal node $n \in Q_{\mathcal{G}}.\varphi$ which has a triple \normalfont{$\langle n, \Rdf{rdf:type}, \Rdf{G:Feature}\rangle$} in $\mathcal{G}$.
\end{definition}

The rationale behind such definition is to ensure that (a) the graph $Q_{\mathcal{G}}.\varphi$ can be safely traversed by joining different sources, and (b) all projected elements are features, which potentially have mappings to the sources. For instance, the SPARQL query depicted in Code \ref{code:Q_nonwellformed}, which retrieves pairs of \Glb{Monitor} and \Glb{FeedbackGathering} per \Glb{SoftwareApplication}, is not well-formed as it retrieves only concepts.  

\begin{lstlisting}[basicstyle=\footnotesize, language=SQL, captionpos=b, label={code:Q_nonwellformed}, mathescape=true, caption=A non well-formed query]
SELECT $?x$, $?y$, $?z$
FROM $\mathcal{G}$
WHERE {
  VALUES ($?x$ $?y$ $?z$) { 
    ($\Rdf{sup:SoftwareApplication}$ $\Rdf{sup:Monitor}$ $\Rdf{sup:FeedbackGathering}$) 
  }
  $\Rdf{sup:SoftwareApplication}$ $\Rdf{sup:hasMonitor}$ $\Rdf{sup:Monitor}$ $\Rdf{.}$
  $\Rdf{sup:SoftwareApplication}$ $\Rdf{sup:hasFGTool}$ $\Rdf{sup:FeedbackGathering}$
}
\end{lstlisting}

In our approach, IDs are considered the default feature. Hence, it is possible to automatically rewrite the query and make it well-formed by replacing projections of concepts for IDs, if available. Such process is depicted in Algorithm \ref{algo:well-formed}, which converts a query to a well-formed one if possible, otherwise it raises an error. Algorithm \ref{algo:well-formed} firstly attempts to detect if the graph pattern $Q_{\mathcal{G}}.\varphi$ is acyclic, which will be true if and only if there exists a topological ordering. Next, it iterates over the projected elements in $Q_{\mathcal{G}}.\pi$ looking for those that are not of type \Rdf{G:Feature} (line 6), in such case it explores all the features of the concept at hand looking for a candidate ID. Note the usage of the auxiliary method $x.\textsc{outgoingNeighborsOfType(}t,g\textsc{)}$, returning, for a node $x$, all outgoing neighbors of type $t$ in the graph $g$ (line 8). Code \ref{code:Q_now_wellformed} depicts the previous non well-formed query now converted to its well-formed version after applying the algorithm.

\begin{algorithm}[h!]\scriptsize
\caption{\label{algo:well-formed}Well-formed query}
\begin{algorithmic}[1]
\Require $\mathcal{T}$ is the BDI ontology, $Q_{\mathcal{G}} = \langle \pi, \varphi \rangle$ is a query over $\mathcal{G}$
\Ensure $Q_{\mathcal{G}}$ is well-formed, otherwise an error is raised
\Function{WellFormedQuery}{$\mathcal{G}, Q_{\mathcal{G}}$}
	\If{$\nexists \textsc{TopologicalSort(}Q_{\mathcal{G}}.\varphi\textsc{)}$}
		\State \Return error($Q_{\mathcal{G}}.\varphi$ has at least one cycle)
	\EndIf
	\ForAll{$\pi \in Q_{\mathcal{G}}.\pi$}
		\If{$\textsc{typeOf(}\pi\textsc{)} \neq \Rdf{G:Feature}$}
			\State $hasID = \texttt{false}$
			\ForAll{$o \in \pi.\textsc{outgoingNeighborsOfType(}\Rdf{"G:Feature"},\mathcal{T}\textsc{)}$}
				\If{$\langle o, \Rdf{"rdfs:subClassOf"}, \Rdf{"sc:identifier"}\rangle \in \mathcal{T}$}
					\State $hasID = \texttt{true}$
					\State $Q_{\mathcal{G}}.\pi = (Q_{\mathcal{G}}.\pi \setminus \{ \pi \}) \cup \{ o \}$
					\State $Q_{\mathcal{G}}.\varphi \ue \langle \pi, \Rdf{"G:hasFeature"}, o \rangle$
				\EndIf
			\EndFor
			\If{$\neg hasID$}
				\State \Return error($Q_{\mathcal{G}}$ has at least one concept without any feature included in the query that is mapped to the sources)
			\EndIf
		\EndIf
	\EndFor
	\State \Return $\mathcal{S}$
\EndFunction
\end{algorithmic}
\end{algorithm}

\begin{lstlisting}[basicstyle=\footnotesize, language=SQL, captionpos=b, label={code:Q_now_wellformed}, caption=A well-formed query, mathescape=true]
SELECT $?x$ $?y$ $?z$
FROM $\mathcal{G}$
WHERE {
  VALUES (?x ?y ?z) {
    ($\Rdf{sup:applicationId}$ $\Rdf{sup:monitorId}$ $\Rdf{sup:feedbackGatheringId}$) 
  }
  $\Rdf{sup:SoftwareApplication}$ $\Rdf{sup:hasMonitor}$ $\Rdf{sup:Monitor}$ $\Rdf{.}$
  $\Rdf{sup:SoftwareApplication}$ $\Rdf{sup:hasFGTool}$ $\Rdf{sup:FeedbackGathering}$ $\Rdf{.}$
  $\Rdf{sup:SoftwareApplication}$ $\Rdf{G:hasFeature}$ $\Rdf{sup:applicationId}$ $\Rdf{.}$
  $\Rdf{sup:Monitor}$ $\Rdf{G:hasFeature}$ $\Rdf{sup:monitorId}$ $\Rdf{.}$
  $\Rdf{sup:FeedbackGathering}$ $\Rdf{G:hasFeature}$ $\Rdf{sup:feedbackGatheringId}$
}
\end{lstlisting}

\subsection{Query rewriting}

The core of the query answering method is the query rewriting algorithm that, given a well-formed query $Q_{\mathcal{G}}$ automatically resolves the LAV mappings and returns a union of conjunctive queries over the wrappers. Intuitively, the algorithm consists of three phases:
\begin{enumerate}
	\item \textit{Query expansion}, which deals with the analysis of the query w.r.t. the ontology. To this end, it takes as input a well-formed query $Q_{\mathcal{G}}$ in order to build its \textit{expanded} version. An expanded query $Q_{\mathcal{G}}^\prime$ contains the same elements as the original $Q_{\mathcal{G}}$, however it also includes IDs for concepts that have not been explicitely requested by the analyst. This is necessary to perform joins in the next phases. In this phase, we also identify which are the concepts in the query, as the next phases are concept-centric.
	\item \textit{Intra-concept generation}, which receives as input the expanded query and generates a list of \textit{partial walks} per concept. Such partial walks indicate how to query the wrappers in order to obtain the requested features for the concept at hand. To achieve this, we utilize SPARQL queries that aid us to obtain the features per concept, as well as to resolve the LAV mappings. 
	\item \textit{Inter-concept generation}, it receives the list of partial walks per concept and joins them to produce covering walks. As result, it returns the union of all the covering and minimal walks found. This is achieved by generating all combinations of partial conjunctive queries that can be joined and that cover the projected attributes in $Q_{\mathcal{G}}$.
\end{enumerate}

Next, we present the algorithms corresponding to each of the phases and their details.

\paragraph{Phase \#1 (query expansion)} The expansion phase (see Algorithm \ref{algo:phase1}) breaks down to the following steps:

\begin{enumerate}[label=\protect\circled{\arabic*}]

\item \textbf{Identify query-related concepts.} The list of query-related concepts consists of vertices of type \Rdf{G:Concept} in the graph pattern (line 4). Traversing $Q_{\mathcal{G}}.\varphi$ we manage to store adjacent concepts in the query in the list $concepts$ (line 5). For the sake of conciseness, algorithms assume linear traversals amongst concepts. Note that using tree-shaped concept traversals is possible, but entails overburdening the algorithms with graph manipulations instead of lists.

\textit{Example.} In the running example (see Figure \ref{fig:queryGlobal}), the list $concepts$ would be $[ \Rdf{sc:SoftwareApplication}$, $\Rdf{sup:Monitor}$, $  \Rdf{sup:InfoMonitor} ]$.

\item \textbf{Expand $Q_{\mathcal{G}}$ with IDs.} Given the list of query-related concepts, we identify their features of type ID by means of a SPARQL query and store it in the set $IDs$ (line 10). For each element in the set $IDs$ we finally expand the query with it (line 12).

\textit{Example.} The expanded query $Q_{\mathcal{G}}^{\prime}$ would include the feature $\Rdf{sup:monitorId}$ (i.e., the ID of concept $\Rdf{sup:Monitor}$), which was not initially in $Q_{\mathcal{G}}$.

\end{enumerate}

\begin{algorithm}[!h]\scriptsize
\caption{\label{algo:phase1} Query Expansion}
\begin{algorithmic}[1]
\Require $Q_{\mathcal{G}}$ is a well-formed query, $\mathcal{T}$ is the BDI ontology
\Ensure $concepts$ is the list of query related concepts, $Q_{\mathcal{G}}^{\prime}$ is the expanded version of $Q_{\mathcal{G}}$ with IDs
\Function{QueryExpansion}{$Q_{\mathcal{G}}, \mathcal{G}$}
	\State $concepts = [\textsc{ }]$
	\For{$v \in \textsc{TopologicalSort(}Q_{\mathcal{G}}.\varphi\textsc{)}$}
		\If{$\langle v, \Rdf{"rdf:type"}, \Rdf{"G:Concept"} \rangle \in \mathcal{T}$}
			\hspace*{17.5em}\rlap{\smash{$\left.\begin{array}{@{}c@{}}\\{}\\{}\\{}\\{}\end{array}\right\}\begin{tabular}{l}\circled{1}\end{tabular}$}}
			\State $concepts.\textsc{add(}v\textsc{)}$
		\EndIf
	\EndFor
	\State $Q_{\mathcal{G}}^{\prime} = Q_{\mathcal{G}}$
	\For{$c \in concepts$}
		\State $IDs = \textrm{SELECT } ?t \textrm{ FROM } \mathcal{T} \textrm{ WHERE } $\par
		\hskip\algorithmicindent $ \{ \langle c,\Rdf{"G:hasFeature"},?t \rangle $.$ \langle ?t$, $ \Rdf{"rdfs:subClassOf"}$, $ \Rdf{"sc:identifier"} \rangle \}$
		\For{$f_{ID} \in IDs$}
		\hspace*{27em}\rlap{\smash{$\left.\begin{array}{@{}c@{}}\\{}\\{}\\{}\\{}\\{}\\{}\end{array}\right\}\begin{tabular}{l}\circled{2}\end{tabular}$}}
			\State $Q_{\mathcal{G}}^{\prime}.\varphi \ue \langle c, \Rdf{"G:hasFeature"}, f_{ID} \rangle$
		\EndFor
	\EndFor
	\State \Return $\langle concepts, Q_{\mathcal{G}}^{\prime} \rangle$
\EndFunction
\end{algorithmic}
\end{algorithm}

\paragraph{Phase \#2 (intra-concept generation)} The intra-concept phase (see Algorithm \ref{algo:phase2}) gets as input the list of concepts in the query, and the expanded query $Q_{\mathcal{G}}^{\prime}$, and outputs the list of partial walks per concept ($partialWalks$ defined in line 2). A partial walk is a walk that is not yet traversing all the concepts required by the query. The process breaks down to the following steps:

\begin{enumerate}[label=\protect\circled{\arabic*}]
\setcounter{enumi}{2}

\item \textbf{Identify queried features.} Phase \#2 starts iterating for each concept in the query. First, we define the auxiliary hashmap $PartialWalksPerWrapper$ (line 5), where its keys are wrappers and its values are walks. To populate this map, we obtain the requested features in $Q_{\mathcal{G}}^{\prime}$ for the concept at hand, which is stored in the set $features$ that is obtained via a SPARQL query over the graph pattern $Q_{\mathcal{G}}^{\prime}.\varphi$ (line 6). 

\textit{Example.} The set $features$ (result of the SPARQL query in line 6) would be $\{ \Rdf{sup:lagRatio}, \Rdf{sup:monitorId}, \Rdf{sup:applicationId} \}$. 

\item \textbf{Unfold LAV mappings.} Next, for each feature $f$ in the set $features$, we look for wrappers whose LAV mapping contain it. This is achieved querying the named graphs in $\mathcal{T}$ (line 8). At this point, we have the information of which wrappers may provide the feature at hand. 

\textit{Example.} For the feature $\Rdf{sup:lagRatio}$ the identified set of wrappers would be $\{ \Rdf{sup:W1} \}$. Likewise, for the feature $\Rdf{sup:monitorId}$ the set $\{ \Rdf{sup:W1}, \Rdf{sup:W3} \}$ and for $\Rdf{sup:applicationId}$ the set $\{ \Rdf{sup:W3} \}$.

\item \textbf{Find attributes in $\mathcal{S}$.} Now, for each wrapper $w$ in the previously devised set of wrappers for feature $f$, with a SPARQL query (line 10) we find the attribute $a$ in $\mathcal{S}$ that maps to the feature at hand (i.e., $\Rdf{owl:sameAs}$ relationship). This will be added to the hashmap $PartialWalksPerWrapper$, with key $w$ and value $\widetilde{\Pi}_{a}(w)$.

\textit{Example.} For feature $\Rdf{sup:lagRatio}$ and wrapper $\Rdf{sup:W1}$, we would identify $\Rdf{sup:D1/lagRatio}$ as attribute in $\mathcal{S}$. Hence, we would add to the hashmap $PartialWalksPerWrapper$ an entry with key $\Rdf{sup:W1}$ and value $\widetilde{\Pi}_{\Rdf{sup:D1/lagRatio}}(\Rdf{sup:W1})$. The process would be likewise for the rest of features and wrappers.

\item \textbf{Prune output.} Note that we might have considered walks that do not contain all the requested features for the current concept $c$ (e.g., a wrapper $w_5$ where $\Src{lagRatio}$ has been dropped), hence, in order to avoid the complexity that combining wrappers within a concept would yield, we only keep those wrappers providing all the features queried for the current concept. To this end, we first use the $\textsc{MergeProjections}$ operator, which merges the projection operators that have been separately added to the walk (e.g., from $\widetilde{\Pi}_{a_1}(w)\widetilde{\Pi}_{a_2}(w)$ to $\widetilde{\Pi}_{a_1,a_2}(w)$). With such wrapper projections, we follow the \Rdf{owl:sameAs} relation from $\mathcal{S}$ to $\mathcal{G}$ to ensure that we are obtaining the same set of features as requested by the analyst (defined in line 6), if so we will add such partial walk to the output, ensuring \textit{covering} and \textit{minimality} for the concept at hand.

\textit{Example.} The final output of phase \#2 would be a list with the following elements: 
\begin{itemize}
	\item $\langle$ \Rdf{sc:SoftwareApplication} $\rightarrow \{ \widetilde{\Pi}_{\Rdf{sup:D3/TargetApp}}(\Rdf{sup:W3}) \} \rangle$
	\item $\langle$ \Rdf{sup:Monitor} $\rightarrow \{ \widetilde{\Pi}_{\Rdf{sup:D1/VoDmonitorId}}(\Rdf{sup:W1}), \widetilde{\Pi}_{\Rdf{sup:D3/MonitorId}}(\Rdf{sup:W3}) \} \rangle$
	\item $\langle$ \Rdf{sup:InfoMonitor} $\rightarrow \{ \widetilde{\Pi}_{\Rdf{sup:D1/lagRatio}}(\Rdf{sup:W1}) \} \rangle$
\end{itemize}
\end{enumerate}

\begin{algorithm}[!h]\scriptsize
\caption{\label{algo:phase2} Intra-concept generation}
\begin{algorithmic}[1]
\Require $concepts$ is the list of concepts in the query, $Q_{\mathcal{G}}^{\prime}$ is an expanded query, $\mathcal{T}$ is the BDI ontology
\Ensure $partialWalks$ is the list of partial walks per concept
\Function{IntraConceptGeneration}{$concepts, Q_{\mathcal{G}}^{\prime}, \mathcal{T}$}
	\State $partialWalks = [\textrm{ }]$
	\For{$i = 0;\textsc{ }i < \textsc{length(}concepts\textsc{)};\textsc{ } \Plus \Plus i$}
		\State $c = concepts[i]$
		\State $PartialWalksPerWrapper = \texttt{HashMap<k,v>}$
		\hspace*{16.5em}\rlap{\smash{$\left.\begin{array}{@{}c@{}}\\{}\\{}\\{}\end{array}\right\}\begin{tabular}{l}\circled{3}\end{tabular}$}}
		\State $features = $ SELECT $?f$ FROM $Q_{\mathcal{G}}^{\prime}.\varphi$ WHERE $\{ \langle c, \Rdf{"G:hasFeature"}, ?f \rangle \}$
		\For{$f \in features$}
			\State $wrappers = \textrm{SELECT } ?g \textrm{ FROM }\mathcal{T} \textrm{ WHERE } $\par
			\hskip\algorithmicindent \tab$ \{ \textrm{ GRAPH } ?g \{ \langle c, \Rdf{"G:hasFeature"}, f \rangle \} \}$
			\hspace*{16.25em}\rlap{\smash{$\left.\begin{array}{@{}c@{}}\\{}\\{}\end{array}\right\}\begin{tabular}{l}\circled{4}\end{tabular}$}}			
			\vspace{.5em}\For{$w \in wrappers$}
				\State $attribute = $ SELECT $?a$ FROM $\mathcal{T}$ WHERE\par
			\hskip\algorithmicindent \tab\tab $\{ \langle ?a, \Rdf{"owl:sameAs"}, f \rangle . \langle w,\Rdf{"S:hasAttribute"}, ?a \rangle \} $
				\hspace*{38em}\rlap{\smash{$\left.\begin{array}{@{}c@{}}\\{}\\{}\\{}\\{}\end{array}\right\}\begin{tabular}{l}\circled{5}\end{tabular}$}}
				\State $PartialWalksPerWrapper[w] \ue \widetilde{\Pi}_{attribute}(w)$
			\EndFor
		\EndFor
		\For{$\langle wrapper,walk \rangle \in PartialWalksPerWrapper$}
			\State $mergedWalk = \textsc{MergeProjections(}walk\textsc{)}$
			\State $featuresInWalk = \{ \}$
			\For{$a \in \textsc{Projections(}mergedWalk\textsc{)}$}
				\State $featuresInWalk\textrm{ }\ue $ SELECT $?f$ FROM $\mathcal{T}$ WHERE\par
			\hskip\algorithmicindent \tab\tab $\{ \langle a, \Rdf{"owl:sameAs"}, ?f \rangle \} $	
			\hspace*{21.125em}\rlap{\smash{$\left.\begin{array}{@{}c@{}}\\{}\\{}\\{}\\{}\\{}\\{}\\{}\\{}\\{}\\{}\\{}\end{array}\right\}\begin{tabular}{l}\circled{6}\end{tabular}$}}
		
			\EndFor
			\If{$featuresInWalk = features$}
				\State $partialWalks.\textsc{add(}\langle c, mergedWalk \rangle \textsc{)}$
			\EndIf
		\EndFor		
	\EndFor
	\State \Return $partialWalks$
\EndFunction
\end{algorithmic}
\end{algorithm}

\paragraph{Phase \#3 (inter-concept generation)} The final phase of the rewriting process (see Algorithm \ref{algo:phase3}) consists of joining the partial walks per concept to obtain a set of walks joining all the concepts required in the query. This is a systematic process where the final list of walks is incrementally built.  

\begin{enumerate}[label=\protect\circled{\arabic*}]
\setcounter{enumi}{6}

\item \textbf{Compute cartesian product.} Phase \#3 iterates on $partialWalks$ using a window of two elements, $current$ (line 2) and $next$ (line 4), and maintain a set of currently joined partial walks (line 5). We start computing the cartesian product of the respective lists of partial walks (line 6), namely $CP_{left}$ (corresponding to $current$) and $CP_{right}$ (corresponding to $next$). 

\textit{Example.} In the first iteration, $current$ and $next$ would be respectively the elements $\langle$ \Rdf{sc:SoftwareApplication} $\rightarrow \{ \widetilde{\Pi}_{\Rdf{sup:D3/TargetApp}}(\Rdf{sup:W3}) \} \rangle$ and $\langle$ \Rdf{sup:Monitor} $\rightarrow \{ \widetilde{\Pi}_{\Rdf{sup:D1/VoDmonitorId}}(\Rdf{sup:W1}), \widetilde{\Pi}_{\Rdf{sup:D3/MonitorId}}(\Rdf{sup:W3}) \} \rangle$. Thus, the resulting cartesian product of the sets of partial walks would be the pair $\langle \widetilde{\Pi}_{\Rdf{sup:D3/TargetApp}}(\Rdf{sup:W3}), \widetilde{\Pi}_{\Rdf{sup:D1/VoDmonitorId}}(\Rdf{sup:W1}) \rangle$ and the pair $\langle \widetilde{\Pi}_{\Rdf{sup:D3/TargetApp}}(\Rdf{sup:W3}), \widetilde{\Pi}_{\Rdf{sup:D3/MonitorId}}(\Rdf{sup:W3}) \rangle$.

\item \textbf{Merge walks.} Given the two partial walks from the cartesian product, the goal is now to merge them into a single one. To this end, we use the function $\textsc{MergeWalks}$ (line 7) that given the two partial walks generates a merged one that projects the attributes from both inputs. At this moment there are two possibilities, (a) there is a wrapper shared by both partial walks and then the join has been materialized by it, or (b) they do not share a wrapper, thus we need to explore ways to join them. In the former case, as discussed, no further join needs to be added to the merged walk, however the latter needs to be extended by an additional join ($\widetilde{\bowtie}$) between both inputs. Such discovery process is described in the following steps.

\textit{Example}. Given $\langle \widetilde{\Pi}_{\Rdf{sup:D3/TargetApp}}(\Rdf{sup:W3}),$ $\widetilde{\Pi}_{\Rdf{D3/MonitorId}}(\Rdf{sup:W3}) \rangle$, the merged walk would be $\widetilde{\Pi}_{\Rdf{sup:D3/TargetApp},\Rdf{sup:D3/MonitorId}}(\Rdf{sup:W3})$ where no extra joins should be added. Regarding the pair $\langle \widetilde{\Pi}_{\Rdf{sup:D3/TargetApp}}(\Rdf{sup:W3}), \widetilde{\Pi}_{\Rdf{sup:D1/VoDmonitorId}}(\Rdf{sup:W1}) \rangle$, after merging the walks the result would be $\widetilde{\Pi}_{\Rdf{sup:D3/TargetApp}}(\Rdf{sup:W3}) \widetilde{\Pi}_{\Rdf{sup:D1/VoDmonitorId}}(\Rdf{sup:W1})$, thus it is necessary to discover how to join $\Rdf{sup:W1}$ and $\Rdf{sup:W3}$.

\item \textbf{Discover join wrappers.} For each pair of concepts related by an edge in $Q_{\mathcal{G}}^{\prime}$ ($current$ and $next$), we aim at retrieving the list of wrappers providing the required features (i.e., identified as partial walks in the previous step). Since $\mathcal{G}$ is a directed graph, we first need to identify, for each edge, the concept playing the role of $current$ and $next$ (e.g., if \Rdf{sc:SoftwareApplication} and \Rdf{sup:Monitor} play the role of $current$ and $next$, respectively, then the join must be computed using the ID of $next$). This is computed in two SPARQL queries (lines 9 and 10). Note that only one direction will be available since our graph query ($Q_{\mathcal{G}}^{\prime}$) does not contain cycles.

\textit{Example.} Given that $current.c$ and $next.c$ are respectively the concepts \Rdf{sc:SoftwareApplication} and \Rdf{sup:Monitor}, as the edge is directed from the former to the latter, only \textit{wrappersFromLtoR} would contain any data, precisely the set of wrappers $\{ \Rdf{sup:W1} \}$. This entails that we need to look for the attribute of type ID for concept \Rdf{sup:Monitor} that is provided by $\Rdf{sup:W1}$.

\item \textbf{Discover join attribute.} Focusing on the case where $next$ must provide the ID (lines 12-17), we start issuing a SPARQL query that tells us such ID (line 12). Next, the operation $\textsc{findWrapperWithID}$ (line 13) identifies which wrapper is providing such ID for $next$, and subsequently we obtain the physical attribute (line 14). Then, we iterate on all wrappers that contribute to the relation between both concepts, and for each wrapper we identify the ID attribute for $left$ (line 16). With such, we can generate a new walk by joining each potential pair resulting from the list of IDs for $current$ and the one identified for $next$ (line 17). As we previously discussed, this process depends on the direction of the edge, therefore line 20 entails that the same process should be executed if the edge goes from $next$ to $current$.

\textit{Example.} Given the partial walks from the previous example, the output of phase \#3 would consist of the following set of walks:

\begin{itemize}
	\item $\widetilde{\Pi}_{\Rdf{sup:D1/lagRatio}, \Rdf{sup:D1/VoDmonitorId}, \Rdf{sup:D3/TargetApp}} \\ \tab (\Rdf{sup:W1} \underset{\Rdf{sup:D1/VoDmonitorId} = \Rdf{sup:D3/MonitorId}}{\widetilde{\bowtie}} \Rdf{sup:W3})$
	\item $\widetilde{\Pi}_{\Rdf{sup:D1/lagRatio}, \Rdf{sup:D3/MonitorId}, \Rdf{sup:D3/TargetApp}} \\ \tab (\Rdf{sup:W1} \underset{\Rdf{sup:D1/VoDmonitorId} = \Rdf{sup:D3/MonitorId}}{\widetilde{\bowtie}} \Rdf{sup:W3})$
\end{itemize}

Note that, even though the analyst requested only the first and third attributes our approach has generated further combinations when considering IDs (in Step 2). Those can be easily projected out at the final step, when generating the union of conjunctive queries.

\end{enumerate}

\begin{algorithm}[h!]\scriptsize
\caption{\label{algo:phase3} Inter-concept generation}
\begin{algorithmic}[1]
\Require $partialWalks$ is the list of partial walks per concept, $\mathcal{S}$ is the source graph and $\mathcal{M}$ the LAV mappings
\Ensure $walks$ is the final list of walks
\Function{InterConceptGeneration}{$partialWalks, \mathcal{S}, \mathcal{M}$}
	\State $current = partialWalks[0]$
	\For{$i = 1;\textsc{ }i < \textsc{length(}partialWalks\textsc{)};\textsc{ } \Plus \Plus i$}
		\State $next = partialWalks[i]$
		\State $joined = \{ \}$
		\For{$\langle CP_{left}, CP_{right} \rangle \in current.lw \times next.lw$}
		\hspace*{14.5em}\rlap{\smash{$\left.\begin{array}{@{}c@{}}\end{array}\right\}\begin{tabular}{l}\circled{7}\end{tabular}$}}
		\vspace{0.5em}		
		\State $mergedWalk = \textsc{mergeWalks(}CP_{left}, CP_{right}\textsc{)}$
		\hspace*{14.875em}\rlap{\smash{$\left.\begin{array}{@{}c@{}}\end{array}\right\}\begin{tabular}{l}\circled{8}\end{tabular}$}}	
		\If{$wrappers(CP_{left}) \cap wrappers(CP_{right}) = \emptyset$} 		
			\State $wrappersFromLtoR = $ SELECT $?g$ FROM $\mathcal{T}$ WHERE\par
			\hskip\algorithmicindent \tab $\{$ GRAPH $?g$ $\{\langle current.c, ?x, next.c \rangle \} \}$
			\hspace*{15.75em}\rlap{\smash{$\left.\begin{array}{@{}c@{}}\\{}\\{}\end{array}\right\}\begin{tabular}{l}\circled{9}\end{tabular}$}}
			\State $wrappersFromRtoL = $ SELECT $?g$ FROM $\mathcal{T}$ WHERE\par
			\hskip\algorithmicindent \tab $\{$ GRAPH $?g$ $\{\langle next.c, ?x, current.c \rangle \} \}$
			\If{$wrappersFromLtoR \neq \emptyset$}
				\State $f_{ID} = $ SELECT $?t$ FROM $\mathcal{T}$ WHERE\par
			\hskip\algorithmicindent \tab\tab $\{ \langle next.c, \Rdf{G:hasFeature}, ?t \rangle $.$ \langle ?t$, $ \Rdf{rdfs:subClassOf}$, $ \Rdf{sc:identifier} \rangle \}$
				\State $wrapperWithID_{right} = \textsc{findWrapperWithID(}CP_{right}\textsc{)}$
				\State $att_{right} = $ SELECT $?a$ FROM $\mathcal{T}$ WHERE\par
			\hskip\algorithmicindent \tab\tab $\{ \langle ?a, \Rdf{owl:sameAs}, f_{ID} \rangle . \langle wrapperWithID_{right},\Rdf{S:hasAttribute}, ?a \rangle \} $
				\hspace*{38.125em}\rlap{\smash{$\left.\begin{array}{@{}c@{}}\\{}\\{}\\{}\\{}\\{}\\{}\\{}\\{}\\{}\end{array}\right\}\begin{tabular}{l}\circled{10}\end{tabular}$}}
				\For{$w \in wrappersFromLtoR$}
					\State $att_{left} = $ SELECT $?a$ FROM $\mathcal{T}$ WHERE\par
			\hskip\algorithmicindent \tab\tab\tab $\{ \langle ?a, \Rdf{owl:sameAs}, f_{ID} \rangle . \langle w,\Rdf{S:hasAttribute}, ?a \rangle \} $
					\State $mergedWalk \ue w \underset{att_{left} = att_{right}}{\widetilde{\bowtie}} wrapperWithID_{right}$
				\EndFor
			\ElsIf{$wrappersFromRtoL \neq \emptyset$}		
				\State Repeat the process from lines 12-17 inverting left and right.
			\EndIf
		\EndIf
		\State $joined.\textsc{add(}mergedWalk\textsc{)}$
		\EndFor
		\State $current = \langle next.c, joined \rangle$
	\EndFor
	\State \Return $current$
\EndFunction
\end{algorithmic}
\end{algorithm}

\subsection{Computational complexity}

The query rewriting algorithm is divided into three blocks, hence we will present the study of the computational complexity for each of them. We will study the complexity in terms of the number of walks generated in the worst case. Such worst case occurs when each concept features is provided by a different wrapper (which forces us to generate more joins) and for each concept different sources provide wrappers for it (which generates unions of alternative walks), which forces us to generate a larger number of joins. 

\begin{itemize}
	\item Phase \#1: this phase expands the query with IDs not explicitly queried and therefore it is linear in the number of concepts in the query.
	\item Phase \#2: this phase is linear in the number wrappers providing all the required features of a given concept of the query. This complexity results from the fact that either a wrapper provides all the features of a concept or it is not considered. Thus, no combinations between wrappers are performed to obtain the features or a given concept. Thus, the output of such phase is an array, where each of its buckets is the size of the number of wrappers per concept ($[(\#W)_{C_1}, (\#W)_{C_2}, \ldots, (\#W)_{C_n}]$).
	\item Phase \#3: this phase yields an exponential complexity as it generates joins of partial walks. Note that a cartesian product is performed for each partial walk of a given concept $c$ in the query. Hence, in the worst case (i.e., all partial walks can be joined), we are generating all combinations of wrappers in order to join them (i.e., $(\#W)_{C_1} \times (\#W)_{C_2} \times \ldots \times (\#W)_{C_n}$).
\end{itemize}

\begin{figure*}[!h]
\centering
\includegraphics[width=1\linewidth]{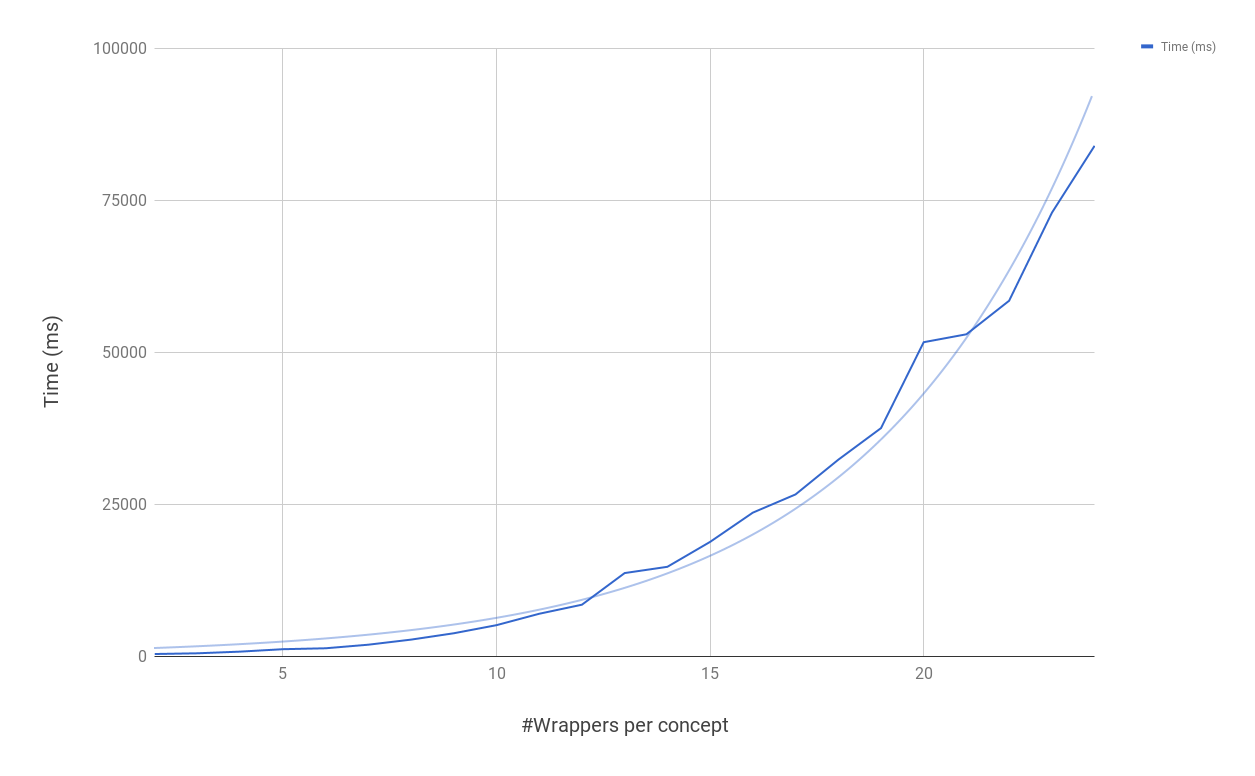}
\caption{Evolution of query answering time in the worst case scenario where wrappers are disjoint (i.e., there is no evolution). The query is a query with 5 concepts. The x-axis shows the number of (disjoint) wrappers per concept.}  \label{fig:times}
\end{figure*}

With the previous discussion, we conclude that in the worst case we can upper bound the theoretical complexity to $\mathcal{O}(W^C)$, assuming each concept has $W$ wrappers generating partial walks (see phase 2), and the query navigates over $C$ concepts. Indeed, such complexity depends on the number of mappings that refer to the query subgraph. To verify the theoretical complexity we have performed a controlled experiment. We have constructed an artificial query navigating through 5 concepts and we have progressively increased the number of wrappers per concept from 1 to 25. Then, we measured the time needed to run the algorithms. This is depicted in Figure \ref{fig:times}, the theoretical prediction (thin line) closely aligns with the observed performance (thick line).
Despite the exponential behavior of query answering, we advocate that realistic Big Data scenarios (e.g., the SUPERSEDE running example) where data are commonly ingested in the form of events, such disjointness in wrappers amongst concepts is not common. In that case, there are few combinations to walk through edges in $\mathcal{G}$, and thus query answering remains tractable in practice.

\section{Evaluation} \label{sec:experiments}

In this section, we present the evaluation results of our approach. We first discuss its implementation, and then provide three kinds of evaluations: a functional evaluation on evolution management, the industrial applicability of our approach and a study on the evolution of the ontology in a real-world API.

\subsection{Implementation}

\begin{figure*}[!h]
\centering
\includegraphics[width=1\linewidth]{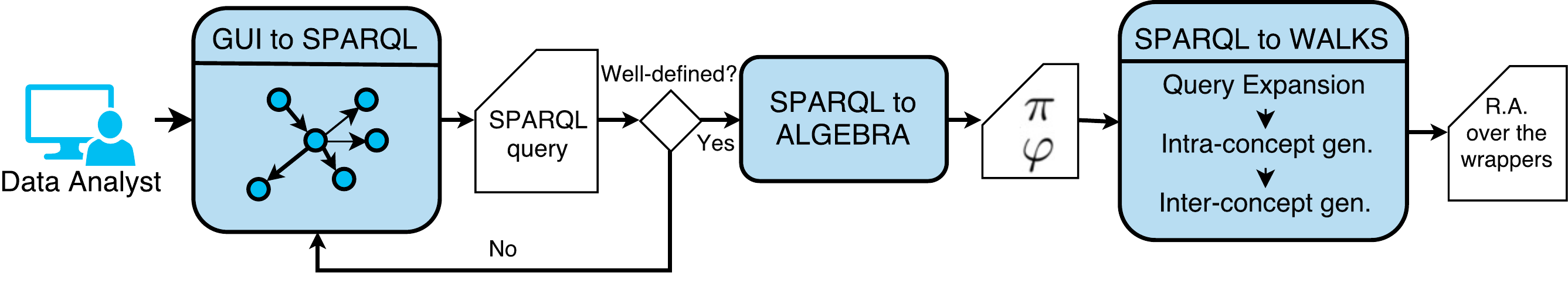}
\caption{Architectural overview of the query answering process} \label{fig:architecture}
\end{figure*}

Prior to discuss the evaluation of our approach we present its implementation, which is part of a system named Metadata Management System (shortly MDM). Figure \ref{fig:architecture} depicts a functional overview of the querying process in the system. Data analysts are presented with a graph-based representation of $\mathcal{G}$ in a user interface where they can graphically pose OMQs. Such graphical representation is automatically converted to its equivalent SPARQL query, and if its well-defined to its algebraic expression $Q_{\mathcal{G}}$. Next, this is the input to our three-phase algorithm for query answering, which will yield a list of walks (i.e., relational algebra expressions over the wrappers).

\begin{figure}[h!]
\centering
\includegraphics[width=1\linewidth]{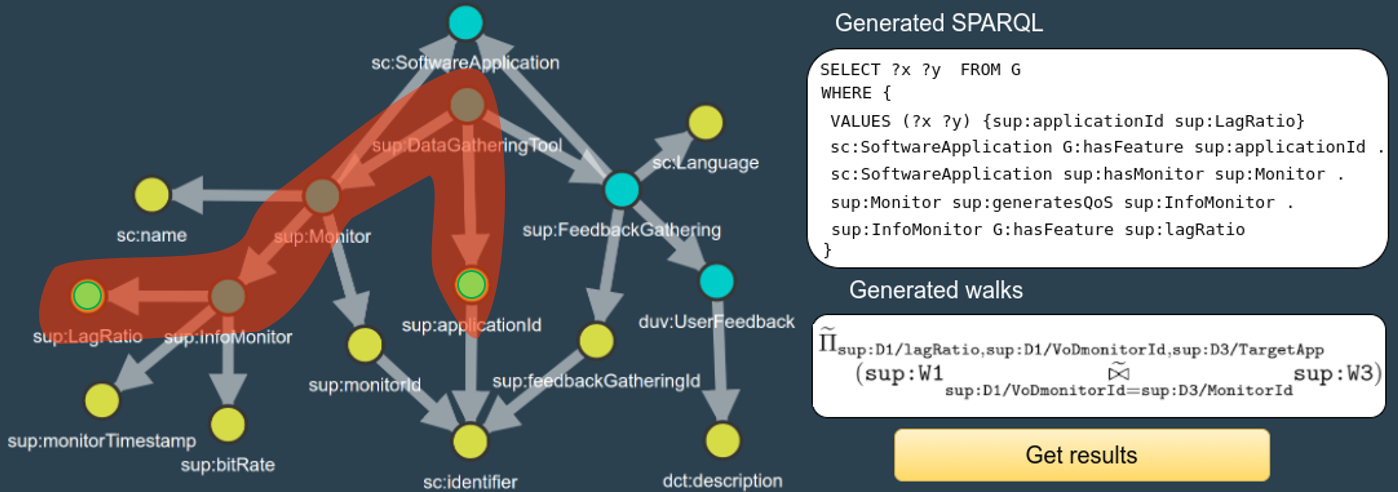}
\caption{Posing an OMQ through the interface and the generated output} \label{fig:screenshot}
\end{figure}

MDM is implemented using a service-oriented architecture. In the frontend, it provides the web-based component to assist the management of the Big Data evolution lifecycle. This component is implemented in JavaScript and resides in a Node.JS web server, Figure \ref{fig:screenshot} depicts an screenshot of the interface to query $\mathcal{G}$. The backend is implemented as a set of REST APIs defined with Jersey for Java. The backend makes heavy use of Jena to deal with RDF graphs, as well as its persistence engine Jena TDB.

\subsection{Functional evaluation}

In order to evaluate the functionalities provided by the BDI ontology, we take the most recent study on structural evolution patterns in REST API \cite{DBLP:conf/icsoc/WangKZ14}. Such work distinguishes changes at 3 different levels, those in (a) API-level, (b) method-level and (c) parameter-level. Our goal is to demostrate that our approach can semi-automatically accommodate such changes. To this end, it is necessary to make a distinction between those changes occurring in the data requests and those in the response. The former are handled by the wrapper's underlying query engine, which also needs to deal with other aspects such as authentication or HTTP query parametrization. The latter will be handled by the proposed ontology.

\paragraph{API-level changes}

Those changes concern the whole of an API. They can be observed either because a new data source is incorporated (e.g., a new social network in the SUPERSEDE use case) or because all methods from a provider have been updated. Table \ref{tab:API_level} depicts the API-level change breakdown and the component responsible to handle it.

\begin{table}[!h]
\centering
\begin{tabular}{|l|c|c|}
\hline
\textbf{API-level Change} & \multicolumn{1}{l|}{\textbf{Wrapper}} & \multicolumn{1}{l|}{\textbf{BDI Ont.}} \\ \hline
Add authentication model          & \tick                                 &                                        \\ \hline
Change resource URL               & \tick                                 &                                        \\ \hline
Change authentication model       & \tick                                 &                                        \\ \hline
Change rate limit                 & \tick                                 &                                        \\ \hline
Delete response format            &                                       & \tick                                  \\ \hline
Add response format               &                                       & \tick                                  \\ \hline
Change response format            &                                       & \tick                                  \\ \hline
\end{tabular}
\caption{API-level changes dealt by wrappers or BDI ontology} \label{tab:API_level}
\end{table}

Adding or changing a response format at API level consists of, for each wrapper querying it, registering a new release with this format. Regarding the deletion of a response format, it does not require actions, due to the fact that no further data on such format will arrive. However, in order to preserve historic backwards compatibility, no elements should be removed from $\mathcal{T}$.

\paragraph{Method-level changes}

Those changes concern modifications on the current version of an operation. They occur either because a new functionality is released or because existing functionalities are modified. Table \ref{tab:method_level} summarizes the method-level change breakdown and the component responsible to handle it.

\begin{table}[!h]
\centering
\begin{tabular}{|l|c|c|}
\hline
\textbf{Method-level Change} & \multicolumn{1}{l|}{\textbf{Wrapper}} & \multicolumn{1}{l|}{\textbf{BDI Ont.}} \\ \hline
Add error code                       & \tick                                 &                                        \\ \hline
Change rate limit                    & \tick                                 &                                        \\ \hline
Change authentication model          & \tick                                 &                                        \\ \hline
Change domain URL                    & \tick                                 &                                        \\ \hline
Add method                           & \tick                                 & \tick                                  \\ \hline
Delete method                        & \tick                                 & \tick                                  \\ \hline
Change method name                   & \tick                                 & \tick                                  \\ \hline
Change response format               &                                       & \tick                                  \\ \hline
\end{tabular}
\caption{Method-level changes dealt by wrappers or BDI ontology} \label{tab:method_level}
\end{table}

Those changes have more overlapping with the wrappers due to the fact that new methods require changes in both request and response. In the context of the BDI ontology, each method is an instance of \texttt{S:DataSource} and thus, adding a new one consists of declaring a new release and running Algorithm \ref{algo:newRelease}. Renaming a method requires renaming the data source instance. As before, a removal does not entail any action with the aim of preserving backwards historic compatibility.

\paragraph{Parameter-level changes}

Such changes are those concerning schema evolution and are the most common on new API releases. Table \ref{tab:parameter_level} depicts such changes and the component in charge of handling it.

\begin{table}[!h]
\centering
\begin{tabular}{|l|c|c|}
\hline
\textbf{Parameter-level Change} & \multicolumn{1}{l|}{\textbf{Wrapper}} & \multicolumn{1}{l|}{\textbf{BDI Ont.}} \\ \hline
Change rate limit               & \tick                                 &                                        \\ \hline
Change require type             & \tick                                 &                                        \\ \hline
Add parameter                   & \tick                                 & \tick                                  \\ \hline
Delete parameter                & \tick                                 & \tick                                  \\ \hline
Rename response parameter       &                                       & \tick                                  \\ \hline
Change format or type           &                                       & \tick                                  \\ \hline
\end{tabular}
\caption{Parameter-level changes dealt by wrappers or BDI ontology} \label{tab:parameter_level}
\end{table}

Similarly to the previous level, some parameter-level changes are managed by both wrappers and the ontology. This is caused by the ambiguity of the change statements, and hence we might consider both URL query parameters and response parameters (i.e., attributes). Changing format of a parameter has a different meaning as before, and here entails a change of data type or structure.
Any of the parameter-level changes identified can be automatically handled by the same process of creating a new release for the source at hand.

\subsection{Industrial applicability}

After functionally validating that the BDI ontology and wrappers can handle all types of API evolution, next we aim to study how these changes occur in real-world APIs. With this purpose, we study the results from \cite{DBLP:conf/icws/LiXLZ13} which presents 16 change patterns that frequently occur in the evolution of 5 widely used APIs (namely \textit{Google Calendar}, \textit{Google Gadgets}, \textit{Amazon MWS}, \textit{Twitter API} and \textit{Sina Weibo}). With such information, we can show the number of changes per API that could be accommodated by the ontology. We summarize the results in Table \ref{tab:proportionPerCompany}. As before, we distinguish between changes concerning (a) the wrappers, (b) the ontology and (c) both wrappers and ontology. This enables us to measure the percentage of changes per API that can be partially accommodated by the ontology (changes also concerning the wrappers) and those fully accommodated (changes only concerning the ontology). Our results show that for all studied APIs, the BDI ontology could, on average, partially accommodate 48.84\% of changes and fully accommodate 22.77\% of changes. In other words, our semi-automatic approach allows to solve on average 71.62\% of changes.

\begin{table*}[h]
\small
	\centering

\resizebox{\linewidth}{!}{	

\begin{tabular}{|l|c|c|c|c|c|}
\hline
\textbf{API Owner} & \textbf{\shortstack{\#Changes\\ Wrapper}} & \textbf{\shortstack{\#Changes\\ Ontology}} & \textbf{\shortstack{\#Changes\\ Wrapper\&Ontology}} & \textbf{\shortstack{Partially\\ Accommodates}} & \textbf{\shortstack{Fully\\ Accommodates}} \\ \hline
Google Calendar    & 0            & 24           & 23                & 48.94\% & 51.06\%                 \\ \hline
Google Gadgets     & 2            & 6            & 30                & 78.95\% & 15.79\%               \\ \hline
Amazon MWS         & 22           & 36           & 14                & 19.44\% & 50\%             \\ \hline
Twitter API        & 27           & 0            & 25                & 48.08\% & 0\%              \\ \hline
Sina Weibo         & 35           & 3            & 56                & 59.57\% & 3.19\%              \\ \hline
\end{tabular}
}
\caption{Number of changes per API and percentage of partially and fully accommodated changes by $\mathcal{T}$} \label{tab:proportionPerCompany}
\end{table*}

\subsection{Ontology evolution}

Now, we are concerned with performance aspects of using the ontology. Particularly, we will study its temporal growth w.r.t. the releases of a real-world API, namely Wordpress REST API\footnote{\url{https://wordpress.org/plugins/rest-api}}. This analysis is of special interest, considering that the size of the ontology may have a direct impact on the cost of querying and maintaining it. As a measure of growth, we count the number of triples in $\mathcal{S}$ after each new release, as it is the most prone to change. Given the high complexity of such APIs, we focus on a specific method and study its structural changes, namely the \textit{GET Posts} API. By studying the changelog, we start from the currently deprecated version 1 evolving it to the next major version release 2. We further introduce 13 minor releases of version 2. (the details of the analysis can be found in \cite{EvolutionWordpress}). We assume that a new wrapper providing all attributes is defined for each release.

\begin{figure*}[!h]
\centering
\includegraphics[width=1\linewidth]{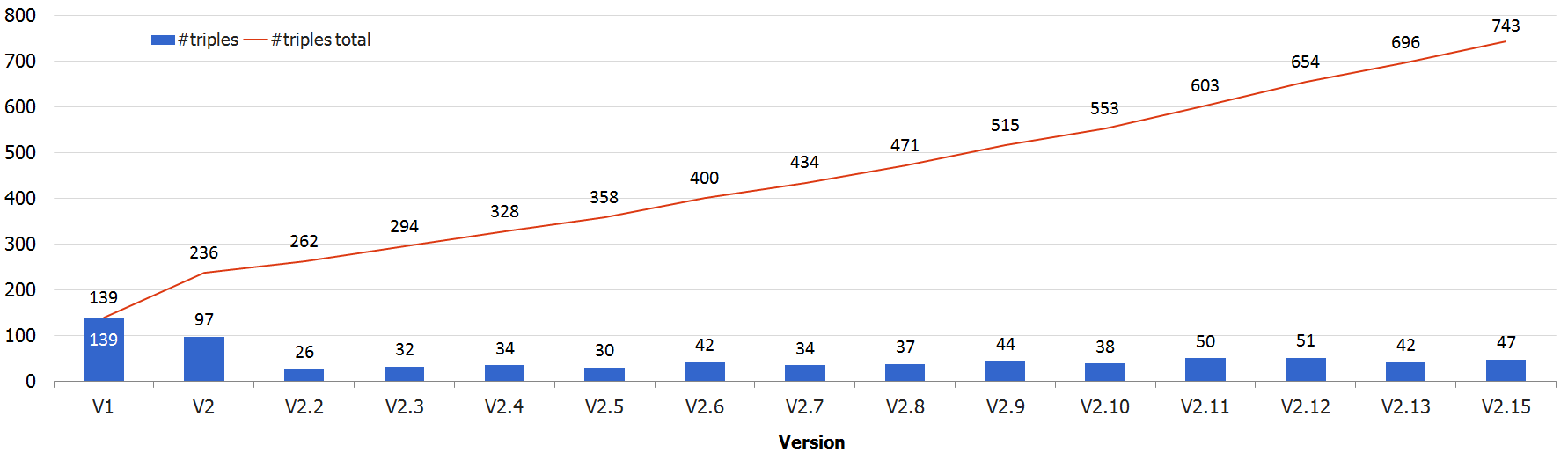}
\caption{Growth in number of triples for $\mathcal{S}$ per release in Wordpress API}  \label{fig:evolutionWordpress}
\end{figure*}

The barcharts in Figure \ref{fig:evolutionWordpress} depict the number of triples added to $\mathcal{S}$ per version release. As version 1 is the first occurrence of such endpoint, all elements must be added and thus carries a big overhead. Version 2 is a major release where few elements can be reused. Later, minor releases do not have many schema changes, with few attribute additions, deletions or renames. Thus, the largest batch of triples per minor release are edges of type $\texttt{S:hasAttribute}$. Each new version needs to identify which attributes it provides even though no change has been applied to it w.r.t. previous versions. 

With such analysis we conclude that major version changes entail a steep growth, however that is infrequent in the studied API. On the other hand, minor versions occur frequently but the growth in terms of triples has a steady linear growth.
The red line depicts the cumulative number of triples after each release. For a practically stable amount of minor release versions, we obtain a linear, stable growth in $\mathcal{S}$. Notice also that $\mathcal{G}$ does not grow. Altogether guarantees that querying $\mathcal{T}$ in query answering will not impose a big overhead, ensuring a good performance of our approach across time. Nonetheless, other optimization techniques (e.g., caching) can be used to further reduce the query cost.



\section{Related work} \label{sec:related}

In previous sections, we have cited relevant works on RESTful API evolution \cite{DBLP:conf/icsoc/WangKZ14,DBLP:conf/icws/LiXLZ13}. They provide a catalog of changes, however they do not provide any approach to systematically deal with them. Other similar works, such as \cite{DBLP:conf/caise/ZarrasVD16}, empirically study API evolution aiming to detect its healthiness.
If we look for approaches that automatically deal with such evolution, we must shift the focus to the area of database schemas, which are mostly focused on relational databases \cite{DBLP:journals/is/SkoulisVZ15,DBLP:journals/jodsn/ManousisVP15}. They apply view cloning to accommodate changes while preserving old views. Such techniques rely on the capability of vetoing certain changes that might affect the overall integrity of the system. This is however an unrealistic approach to adopt in our setting, as schema changes are done by third party data providers.

Attention has also been paid to change management in the context of description logics (DLs). The definition of a DL that provides expresiveness to represent temporal changes in the ontology has been an interesting topic of study in the past years \cite{DBLP:conf/time/LutzWZ08}. Relevant examples include \cite{DBLP:conf/ijcai/ArtaleKWZ13}, that defines the temporal DL \textit{TQL}, providing temporal aspects at the conceptual model level, or \cite{DBLP:conf/apccm/KeetO15} that delves on how to provide such temporal aspects for specific attributes in a conceptual model. It is known, however, that providing such temporal aspects to DLs entails a poor computational behaviour for CQ answering \cite{DBLP:conf/time/LutzWZ08}, for instance the previous examples are respectively coNP-\textit{hard} and undecidable.
Recent efforts are being put to overcome such issues and to provide tractable DLs and methods for rewritability of OMQs. For instance, \cite{DBLP:conf/ijcai/ArtaleKKRWZ15} provides a temporal DL where the cost of first-order rewritability is polynomial, however that is only applicable for a restricted fragment of \textit{DL-Lite}, and besides the notion of temporal attribute, which is key for management of schema evolution does not exist. Generally speaking, most of this approaches lack key characteristics for the management of schema evolution \cite{DBLP:journals/kais/NoyK04}.

Regarding LAV schema mappings in data integration, few approaches strictly follow its definition. This is mostly due to the inherent complexity of query answering in LAV, which is reduced to the problem of answering queries using views \cite{DBLP:conf/pods/LevyMSS95}. Probably the most prominent data integration system that follows the LAV approach is Information Manifold \cite{kirk1995information}. To overcome the complexity posed by LAV query answering, combined approaches of GAV and LAV have been proposed, which are commonly referred as \textit{both-as-view} (BAV) \cite{DBLP:conf/icde/McBrienP03} or \textit{global-and-local-as-view} (GLAV) \cite{DBLP:conf/ijcai/FriedmanLM99}. Oppositely, we are capable of adopting a purely LAV approach by restricting the kind of allowed queries as well as how the mediated schema (i.e., ontology) has to be constructed.

\paragraph{Novelty with respect to the state of the art}
Going beyond the related literature on management of schema evolution, our DOLAP'17 paper \cite{DBLP:conf/edbt/Nadal0AVV17} proposed an RDF vocabulary-based approach to tackle such kind of evolution. Precisely, we focused on Big Data ecosystems that ingest data from REST APIs in JSON format. This paper extends our prior work, where, in the line of the mediator/wrapper architecture, we delegate the complexity of querying the sources to the wrappers. With such, we achieve the possibility to define LAV mappings, which are required in our setting. More importantly, we provide a tractable query answering algorithm that does not require reasoning to resolve LAV mappings.
\section{Conclusions and Future Work} \label{sec:conclusions}

Our research aims at providing self-adapting capabilities in the presence of evolution in Big Data ecosystems. In this paper, we have presented the building blocks to handle schema evolution using a vocabulary-based approach to OBDA. Thus, unlike current OBDA approaches, we restrict the language from generic knowledge representation ontology languages (such as DL-Lite) to ontologies based on RDF vocabularies. We also restrict reasoning to the RDFS entailment regime. These decisions are made to enable LAV mappings instead of GAV. The proposed Big Data integration ontology aims to provide data analysts with an RDF-based conceptual model of the domain of interest, with the limitations that features cannot be reused among concepts. Data sources are accessed via wrappers, which must expose a relational schema in order to depict its RDF-based representation in the ontology and define LAV mappings, by means of named graphs and links from attributes to features. We have defined a query answering algorithm that leverages the proposed ontology and translates a restricted subset of SPARQL queries (see Section \ref{sec:notation}) over the ontology to queries over the sources (i.e., relational expressions on top of the wrappers). Also, we have presented an algorithm to aid data stewards to systematically accommodate announced changes in the form of releases. Our evaluation results show that a great number of changes performed in real-world APIs could be semi-automatically handled by the wrappers and the ontology. We additionally have shown the feasability of our query answering algorithm.
There are many interesting future directions. A prominent one is to extend the ontology with richer constructs to semi-automatically adapt to unanticipated schema changes.

\section*{Acknowledgements}
We thank the reviewers of both this paper and of its earlier version for their constructive comments that have significantly improved the quality of the paper. This work was partly supported by the H2020 SUPERSEDE project, funded by the EU Information and Communication Technologies Programme under grant agreement no 644018, and the GENESIS project, funded by the Spanish Ministerio de Ciencia e Innovación under project TIN2016-79269-R.

\section{References}
\bibliographystyle{abbrv}
\bibliography{references}

\end{document}